\documentclass[10pt,letterpaper]{article}
\usepackage{geometry}
\usepackage{amsmath,amssymb,slashed,bbm}
\usepackage{setspace}

\csname@addtoreset\endcsname{equation}{section}

\makeatletter
\pdfoutput=1

\usepackage[T1]{fontenc}
\usepackage{ae}
\usepackage{aecompl}
\usepackage{soul}

\usepackage[english]{babel}

\usepackage[sf,bf,small]{titlesec}

\usepackage[small,sf,bf,hang]{caption}

\usepackage[multiple]{footmisc}
\long\def\symbolfootnote[#1]#2{\begingroup%
\def\thefootnote{\fnsymbol{footnote}}\footnote[#1]{#2}\endgroup}

\usepackage{titletoc}
\dottedcontents{section}[6mm]{\smallskip }{6mm}{0pc} 
\dottedcontents{subsection}[8mm]{\itshape\small }{6mm}{0pc}
\dottedcontents{subsubsection}[16mm]{\itshape\small }{10mm}{0pc}
\def\tableofcontents{\subsection*{\contentsname}\vspace{-2mm}\@starttoc{toc}}

\usepackage[nosort]{cite} 

\usepackage[parsep]{collref}  
\usepackage{color}

\usepackage{graphicx}
\usepackage{feynmp}  

\DeclareGraphicsExtensions{.pdf}  % include pdf files by default
\def \bea  {\begin{eqnarray}}
\def \eea  {\end{eqnarray}}

\newcommand{\ket}[1]{|{#1}\rangle}
\newcommand{\nn}{\nonumber}
\newcommand{\3}{$AdS_3\times S^3\times S^3\times S^1$}

\newcommand{\4}{$AdS_3\times S^3\times T^4$}

\begin{document}
%%
%% This file contains descriptions of feynman diagrams for use with the package feynMP
%% After running the main file, run "mpost diagrams" once to generate these
%%
%%
%% These are for propagators paper with Per. 
%% The version in this folder edited 18 May 2011 to add unlabled ones for introduction. 
%%

\newsavebox{\feynmanrules}
\sbox{\feynmanrules}{
\begin{fmffile}{diagrams} % I can't seem to make this work using any path but the same one as the document

%%%%%%%%%%%%%%%%
%%  SETTINGS

\fmfset{thin}{0.6pt}  % was 0.7 until v24
%\fmfset{wiggly_len}{5mm}
\fmfset{dash_len}{4pt}
\fmfset{dot_size}{1thick}
\fmfset{arrow_len}{6pt} % you can't use em here, mpost doesn't know what it will be.
%\fmfset{curly_len}{2.5mm}
%\setlength{\unitlength}{1em} % default is =1pt, maybe that's sensible. 72pt = 1in

%%%%%%%%%%%%%%%%%%%%%%%%%%% AdS3 - Scatterings %%%%%%%%%%%%%%%%%%%%%%%%%%%%%%%

%% TREE LEVEL SCATTERING

\begin{fmfgraph*}(50,25)
\fmfkeep{S-t-chan}
\fmfbottom{i1,d1,o1}
\fmftop{i2,d2,o2}
\fmf{fermion}{i1,v1,o1}
\fmf{fermion}{i2,v2,o2}
\fmf{plain,tension=0}{v1,v2}
\fmflabel{$(t)$}{v1}
\fmflabel{$2$}{i1}
\fmflabel{$1$}{i2}
\fmflabel{$4$}{o1}
\fmflabel{$3$}{o2}
\end{fmfgraph*}

\begin{fmfgraph*}(50,25)
\fmfkeep{S-u-chan}
\fmfbottom{i1,d1,o1}
\fmftop{i2,d2,o2}
\fmf{fermion,tension=2}{i1,v1}
\fmf{phantom,tension=1.5}{v1,o1}
\fmf{plain}{v1,v3}
\fmf{fermion}{v3,o2}
\fmf{fermion,tension=2}{i2,v2}
\fmf{phantom,tension=1.5}{v2,o2}
\fmf{plain}{v2,v3}
\fmf{fermion}{v3,o1}
\fmf{plain,tension=-0.5}{v1,v2}
\fmflabel{$(u)$}{v1}
\fmflabel{$2$}{i1}
\fmflabel{$1$}{i2}
\fmflabel{$4$}{o1}
\fmflabel{$3$}{o2}
\end{fmfgraph*}

\begin{fmfgraph*}(50,25)
\fmfkeep{T-t-chan}
\fmfbottom{i1,d1,o1}
\fmftop{i2,d2,o2}
\fmf{fermion}{o1,v1,i1}
\fmf{fermion}{i2,v2,o2}
\fmf{plain,tension=0}{v1,v2}
\fmflabel{$(t)$}{v1}
\fmflabel{$2$}{i1}
\fmflabel{$1$}{i2}
\fmflabel{$4$}{o1}
\fmflabel{$3$}{o2}
\end{fmfgraph*}

\begin{fmfgraph*}(50,25)
\fmfkeep{T-s-chan}
\fmfleft{i1,i2}
\fmfright{o1,o2}
\fmf{fermion}{i2,v1,i1}
\fmf{fermion}{o1,v2,o2}
\fmf{plain,label=$(s)$,label.dist=16}{v1,v2}
\fmflabel{$2$}{i1}
\fmflabel{$1$}{i2}
\fmflabel{$4$}{o1}
\fmflabel{$3$}{o2}
\end{fmfgraph*}

\begin{fmfgraph*}(50,25)
\fmfkeep{R-u-chan}
\fmfbottom{i1,d1,o1}
\fmftop{i2,d2,o2}
\fmf{fermion,tension=2}{i1,v1}
\fmf{phantom,tension=1.5}{v1,o1}
\fmf{plain}{v1,v3}
\fmf{fermion}{o2,v3}
\fmf{fermion,tension=2}{i2,v2}
\fmf{phantom,tension=1.5}{v2,o2}
\fmf{plain}{v2,v3}
\fmf{fermion}{v3,o1}
\fmf{plain,tension=-0.5}{v1,v2}
\fmflabel{$(t)$}{v1}
\fmflabel{$2$}{i1}
\fmflabel{$1$}{i2}
\fmflabel{$4$}{o1}
\fmflabel{$3$}{o2}
\end{fmfgraph*}

\begin{fmfgraph*}(50,25)
\fmfkeep{R-s-chan}
\fmfleft{i1,i2}
\fmfright{o1,o2}
\fmf{fermion}{i2,v1}
\fmf{fermion}{i1,v1}
\fmf{fermion}{v2,o1}
\fmf{fermion}{o2,v2}
\fmf{plain,label=$(s)$,label.dist=16}{v1,v2}
\fmflabel{$2$}{i1}
\fmflabel{$1$}{i2}
\fmflabel{$4$}{o1}
\fmflabel{$3$}{o2}
\end{fmfgraph*}

\begin{fmfgraph*}(50,25)
\fmfkeep{S-c}
\fmfbottom{i1,o1}
\fmftop{i2,o2}
\fmf{fermion}{i1,v1,o1}
\fmf{fermion}{i2,v1,o2}
%\fmf{plain,label=$(s)$}{v1}
%\fmf{fermion,tension=0}{v1,v2}
\fmf{phantom,label=$(c)$}{i1,o1}
\fmflabel{$(t)$}{i2}
\fmflabel{$2$}{i1}
\fmflabel{$1$}{i2}
\fmflabel{$4$}{o1}
\fmflabel{$3$}{o2}
\end{fmfgraph*}

\begin{fmfgraph*}(50,25)
\fmfkeep{T-c}
\fmfbottom{i1,o1}
\fmftop{i2,o2}
\fmf{fermion}{i2,v1,i1}
\fmf{fermion}{o1,v1,o2}
\fmf{phantom,label=$(c)$}{i1,o1}
%\fmf{fermion,tension=0}{v1,v2}
%\fmflabel{$(t)$}{v1}
\fmflabel{$2$}{i1}
\fmflabel{$1$}{i2}
\fmflabel{$4$}{o1}
\fmflabel{$3$}{o2}
\end{fmfgraph*}

\begin{fmfgraph*}(50,25)
\fmfkeep{R-c}
\fmfbottom{i1,o1}
\fmftop{i2,o2}
\fmf{fermion}{i2,v1,i1}
\fmf{fermion}{o1,v1,o2}
\fmf{phantom,label=$(c)$}{i1,o1}
%\fmf{fermion,tension=0}{v1,v2}
%\fmflabel{$(t)$}{v1}
\fmflabel{$2$}{i1}
\fmflabel{$1$}{i2}
\fmflabel{$3$}{o1}
\fmflabel{$4$}{o2}
\end{fmfgraph*}
\begin{fmfgraph*}(50,25)

\fmfkeep{R-t-chan}
\fmfbottom{i1,d1,o1}
\fmftop{i2,d2,o2}
\fmf{fermion}{o1,v1,i1}
\fmf{fermion}{i2,v2,o2}
\fmf{plain,tension=0}{v1,v2}
\fmflabel{$(t)$}{v1}
\fmflabel{$2$}{i1}
\fmflabel{$1$}{i2}
\fmflabel{$3$}{o1}
\fmflabel{$4$}{o2}
\end{fmfgraph*}

\end{fmffile}
}

%\title{Worldsheet scattering in $AdS_3\times S^3\times S^3\times S^1$}

\begin{flushright}
MIFPA-13-08\bigskip\bigskip
\par\end{flushright}

\begin{center}
\textsf{\textbf{\Large Worldsheet scattering in AdS$_3$/CFT$_2$\smallskip\smallskip }}\\
\textsf{\textbf{\Large  }}
\par\end{center}{\Large \par}

\begin{singlespace}
\begin{center}
Per Sundin$^{1}$ and  Linus Wulff$^{2}$ \bigskip \\

{\small $^{1}$}\emph{\small{} The Laboratory for Quantum Gravity \& Strings}\\
\emph{\small Department of Mathematics and Applied Mathematics, }\\
\emph{\small University of Cape Town,}\\
\emph{\small Private Bag, Rondebosch, 7700, South Africa}{\small }\\
\emph{\small nidnus.rep@gmail.com}\vspace{0.2cm}

{\small $^{2}$}\emph{\small{} George P. \& Cynthia Woods Mitchell Institute for Fundamental Physics and Astronomy,}\\
\emph{\small Texas A\&M University, College Station, }\\
\emph{\small TX 77843, USA}\\
\emph{\small linus@physics.tamu.edu }{\small \bigskip }\emph{\small }\\
\par\end{center}{\small \par}
\end{singlespace}

%\begin{center}
%Draft printed \bigskip  \today , from file \jobname.
%\par\end{center}
\bigskip\bigskip

\subsection*{\hspace{9mm}Abstract}
\begin{quote}
%\maketitle
%\begin{abstract}
We confront the recently proposed exact S-matrices for $AdS_3/CFT_2$ with direct worldsheet calculations. Utilizing the BMN and Near Flat Space (NFS) expansions for strings on \3 and $AdS_3\times S^3\times T^4$ we compute both tree-level and one-loop scattering amplitudes. Up to some minor issues we find nice agreement in the tree-level sector. At the one-loop level however we find that certain non-zero tree-level processes, which are not visible in the exact solution, contribute, via the optical theorem, and give an apparent mismatch for certain amplitudes. Furthermore we find that a proposed one-loop modification of the dressing phase correctly reproduces the worldsheet calculation while the standard Hernandez-Lopez phase does not. We also compute several massless to massless processes. 
%\end{abstract}
\bigskip  
\thispagestyle{empty} 
\end{quote}

\newpage
\tableofcontents
\setcounter{page}{1}

\section{Introduction}
The $AdS_3/CFT_2$-correspondence, where the gravity side consists of superstrings on either \4 or \3 supported by pure Ramond-Ramond (RR) flux and preserving 16 supercharges, has recently been seen to have an interesting integrable structure \cite{Babichenko:2009dk,OhlssonSax:2011ms,Forini:2012bb,Rughoonauth:2012qd,Sundin:2012gc,Sax:2012jv,Ahn:2012hw,Borsato:2012ud,Abbott:2012dd,Beccaria:2012kb,Borsato:2012ss,Beccaria:2012pm}\footnote{For the case of both NSNS and RR flux see \cite{Cagnazzo:2012se}.} similar to that of strings in $AdS_5\times S^5$ and $AdS_4\times\mathbbm{CP}^3$. In the more general $S^3\times S^1$ case very little is known about the dual CFT \cite{deBoer:1999rh,Gukov:2004ym} but one can still learn a lot by analyzing the integrable structure of the theory. On the string side one can use the standard approach of a supercoset model description, to formulate sets of quantum Bethe equations \cite{Babichenko:2009dk,Bena:2003wd, Kazakov:2004qf,Staudacher:2004tk,Beisert:2005bm,Beisert:2005tm} (see also \cite{Sundin:2012gc} for a discussion of integrability without fixing the kappa-symmetry of the Green-Schwarz string). Alternatively one can start from the superisometry group of the background and construct a spin-chain from which one can read off the S-matrix (which in turns is diagonalized by the Bethe ansatz equations) \cite{David:2008yk,David:2010yg,Sax:2012jv,Ahn:2012hw,Borsato:2012ud,Borsato:2012ss}. The relevant algebra for \3 is $d(2,1;\alpha)^2$ (this corresponds to a subgroup of the superisometry group of the background $D(2,1;\alpha)^2\times U(1)$). The parameter $0\leq\alpha\leq1$ sets the ratio of the two $S^3$ radii \cite{Babichenko:2009dk}. The $T^4$ corresponds to $\alpha=0,1$ and in this case the relevant algebra becomes $psu(1,1|2)^2=d(2,1;0)^2=d(2,1;1)^2$. 

In this paper we continue the investigation initiated in \cite{Rughoonauth:2012qd, Sundin:2012gc} in order to confront the proposed S-matrices of \cite{Ahn:2012hw} and \cite{Borsato:2012ud,Borsato:2012ss} with explicit worldsheet calculations. We do this both at the tree and one-loop level utilizing the BMN and NFS expansion of the string in \3 and \4 \cite{Berenstein:2002jq,Maldacena:2006rv}, see also \cite{Abbott:2011xp}. The worldsheet excitations consist of four pairs of modes with masses $m=(1,\alpha,1-\alpha,0)$ and at tree-level we compute all possible processes for two light ($m=\alpha,1-\alpha$) incoming bosons. For the case of $AdS_3\times S^3\times T^4$, where we have a four fold degeneracy of heavy ($m=1$) and massless modes, we also perform a one-loop computation in the near flat space (NFS) limit. Upon natural choices of free parameters and some minor adjustments we find that the tree-level sector of the {S-matrix found in \cite{Borsato:2012ud,Borsato:2012ss} reproduces our findings, but \cite{Ahn:2012hw} fails to do so in the sector that involves fermions}. However, we also find that there are non-zero tree-level amplitudes of the form of two light bosons going to two heavy fermions not accounted for in \cite{Borsato:2012ud,Borsato:2012ss} which result in an apparent mismatch at the one-loop level (for technical reasons we only check this in the strict $\alpha=1$ limit). This can be understood in terms of the tree-level amplitudes via the optical theorem, and we explicitly show that the additional missing amplitude accounts for the mismatch. This apparent mismatch may be due to the $\alpha\rightarrow0,1$ limit of {\cite{Borsato:2012ud,Borsato:2012ss} being subtle, but from the worldsheet perspective one would expect the limit to be smooth}. {At the one-loop level we also find that the two modified phases of \cite{Beccaria:2012kb} (one of the phases was first derived in \cite{David:2010yg} and agrees at least to lowest order) reproduce} the correct answer while the standard Hernandez-Lopez phase \cite{Hernandez:2006tk} does not, see also \cite{Abbott:2012dd}.

The outline of the paper is as follows: In section \ref{sec:action-bmn-limit} we review the derivation of the gauge-fixed BMN Lagrangian up to quadratic order in fermions \cite{Rughoonauth:2012qd}. In section \ref{sec:tree-level-scattering} we compute tree-level amplitudes for two incoming light-bosons. First we generalize the light to light $BB\rightarrow BB$ processes in \cite{Rughoonauth:2012qd} to the more general $a$-gauge, which turns out to be useful in comparing to the exact S-matrices of \cite{Ahn:2012hw,Borsato:2012ss}. Then we compute additional tree-level processes in the uniform gauge ($a=\frac12$) where we, in particular, find certain non-zero amplitudes not visible in the exact S-matrix. In section \ref{sec:one-loop-scattering} we utilize the NFS string in the strict $\alpha=1$ limit and compute one-loop amplitudes for processes with massive and massless in / out-states. In section \ref{sec:exact-s-matris} we compare out findings with the proposals of \cite{Ahn:2012hw,Borsato:2012ss}. We first explain how to match our tree-level results with the proposals using some natural choices of the free parameters in \cite{Borsato:2012ss} together with certain gauge-dependent phases. For the one-loop amplitudes we show that the real part of the S-matrix element fails to match with either of the proposals for the exact S-matrix in the $\alpha\rightarrow1$ limit. Using the optical theorem we demonstrate that this mismatch is due to the contribution of the additional tree-level amplitude not visible in the exact S-matrix. We end the paper with a short discussion and outlook and some appendices with details of the calculations.   

\section{Action and BMN limit}
\label{sec:action-bmn-limit}
\subsection{Green-Schwarz action to quadratic order in fermions}
The type IIA Green-Schwarz string action on $AdS_3\times S^3\times S^3\times S^1$, to quadratic order in fermions, takes the form 
\cite{Rughoonauth:2012qd}\footnote{The IIB case considered in \cite{Babichenko:2009dk} is easily obtained by T-dualizing along the $S^1$.}
\begin{equation}
\label{action}
S=-g\int\left(\frac{1}{2}*e^Ae_A+i*e^A\,\Theta\Gamma_A{\mathcal D}\Theta-ie^A\,\Theta\Gamma_A\Gamma_{11}{\mathcal D}\Theta\right)+\mathcal O(\Theta^4)\,,
\end{equation}
where the $e^A(X)$ $(A=0,1,\cdots,9)$ are worldsheet pullbacks of the vielbein one-forms (of the purely bosonic part) of the background ($*$ denotes the worldsheet Hodge-dual and we leave the wedge product implicit), and $\Theta$ is a 32-component Majorana Spinor. The generalized covariant derivative acting on $\Theta$ is given by
\begin{equation}
\label{E}
{\mathcal D}\Theta=(\nabla-\frac{1}{2}e^A\,\Gamma^{0129}(1-\mathcal P)\Gamma_A)\ \Theta\quad\mbox{where}\quad \nabla\Theta=(d-\frac{1}{4}\omega^{AB}\Gamma_{AB})\Theta\,,
\end{equation}
where $\omega^{AB}$ is the spin connection of the background space-time. Here $\mathcal P$ is a projection matrix given by
\begin{equation}
\label{calP}
\mathcal P=\frac{1}{2}(1+\sqrt\alpha\,\Gamma^{012345}+\sqrt{1-\alpha}\,\Gamma^{012678})
\end{equation}
and is in fact the projector which singles out the 16 supersymmetries preserved by the background. The parameter $0\leq\alpha\leq1$ determines the relative size of the two $S^3$'s. For unit $AdS$-radius the $S^3$-radii are
\begin{equation}
\label{eq:radii}
R_+=\frac{1}{\sqrt\alpha}\,,\qquad R_-=\frac{1}{\sqrt{1-\alpha}}
\end{equation}
The case $\alpha=0,1$ corresponds to $AdS_3\times S^3\times T^4$ where one of the three-spheres is decompactified.

\subsection{Gauge fixed BMN Lagrangian}
We will use the coordinates described in Appendix A of \cite{Rughoonauth:2012qd} and we consider a string moving along a geodesic involving the angles $\varphi_5$ and $\varphi_8$ of the two three-spheres, again as in \cite{Rughoonauth:2012qd}. In order to allow for a general light-cone gauge fixing we choose the following light-cone coordinates
\bea 
&&x^+=(1-a)t+a(\sqrt\alpha\,\varphi_5+\sqrt{1-\alpha}\,\varphi_8)\,,\quad x^-=\frac12(t-(\sqrt\alpha\,\varphi_5+\sqrt{1-\alpha}\,\varphi_8))\,,\nn\\
&&t=x^+ + 2a\,x^-\,,\quad v=\sqrt{1-\alpha}\,\varphi_5-\sqrt\alpha\,\varphi_8\,,\\
&&\varphi_5=\sqrt\alpha\big( x^+-2(1-a)x^-\big)+\sqrt{1-\alpha}\,v\,,\quad \varphi_8=\sqrt{1-\alpha}\big( x^+-2(1-a)x^-\big)-\sqrt{\alpha}\,v\nn
\eea 
together with
\bea \nn 
\Gamma_+=\Gamma_0+\sqrt \alpha\,\Gamma_5+\sqrt{1-\alpha}\,\Gamma_8\,,\qquad 
\Gamma_-=2(1-a)\Gamma_0-2a\big(\sqrt\alpha\Gamma_5+\sqrt{1-\alpha}\Gamma_8\big)\,,
\eea 
where $0 \leq a \leq 1$ and generalizes the standard uniform light-cone gauge, see \cite{Arutyunov:2009ga} for a detailed review. The light-cone gauge is then fixed by setting 
\bea \label{eq-gauge}
x^+=\tau,\qquad p^+=1\,,
\eea 
where $p^+$ is the conjugate momentum to $x^-$. The full Lagrangian simplifies considerably in the uniform light-cone gauge where $a=\frac{1}{2}$ and we will mostly restrict to this choice. 

Fixing the gauge and expanding the action (\ref{action}) in transverse fields, with $\frac{\sqrt{\lambda}}{p^+}$ kept fixed, gives the BMN Lagrangian
\bea \nn 
\mathcal{L}=\mathcal{L}_2+g^{-1/2}\mathcal{L}_3+g^{-1}\mathcal{L}_4+...
\eea 
where $g\sim \sqrt{\lambda}$ and will be left implicit in most of the calculations. The action describes four complex bosons, $y_1=\frac{1}{\sqrt2}(x_1-ix_2)$ etc, and complex fermions, $\chi^i_\pm$ $i=1,\ldots,4$. The quadratic part is given by \cite{Rughoonauth:2012qd} ($\partial_\pm=\partial_0\pm \partial_1$)
\bea 
\label{bmncoordinateL}
\mathcal{L}_2= i\bar\chi_+^i\partial_- \chi_+^i+i\bar\chi_-^i\partial_+\chi_-^i
+\frac{1}{2}\partial_+ y_i \partial_-\bar{y}_i
+\frac{1}{2}\partial_- y_i \partial_+\bar{y}_i
-m_i^2 y_i\bar{y}_i
-m_i\big(\bar\chi_+^i \chi_-^i+\bar\chi_-^i\chi_+^i\big)\,,
\eea 
where the masses are 
\begin{equation}
\label{eq:spectrum}
m_1=1,\qquad m_2=\alpha,\qquad m_3=1-\alpha,\qquad m_4=0
\end{equation}
and $\alpha$ parameterizes the relative size of the $S^3$ radii according to (\ref{eq:radii}).\footnote{The relation between $\alpha$ and the angle $\phi$ used in \cite{Rughoonauth:2012qd, Sundin:2012gc} is $\alpha=\cos^2\phi$.}

Thus we have four (complex) coordinates with generally distinct masses. Their charges under the various $U(1)$'s can be read of in table \ref{tab:charges}. 
\begin{table}[h]
\centering
\begin{tabular}{c|c|c|c|c|c|c|c|c}
 & $y_1$ & $y_2$& $ y_3$ & $y_4$ & $\chi_\pm^1$ & $\chi_\pm^2$ & $\chi_\pm^3$& $ \chi_\pm^4$ \\ \hline 
$U(1)_+$ & 0 & 1 & 0 & 0 & 1/2 & 1/2 & 1/2 & 1/2 \\ \hline 
$U(1)_-$ & 0 & 0 & 1 & 0 & 1/2 & -1/2 & -1/2 & 1/2  \\ \hline 
$U(1)_{AdS}$ & 1 & 0 & 0 & 0 & -1/2 & -1/2 & 1/2 & 1/2 
\end{tabular}
\caption{U(1) charges}
\label{tab:charges}
\end{table}
Note that the massless boson $y_4$ has zero charge. 

The cubic Lagrangian takes the form \cite{Rughoonauth:2012qd}\footnote{\label{footnote:switch}Note that we are using a slightly asymmetric notation for the 2 and 3 fermions. The action should be invariant under $2\leftrightarrow 3$ together with $\alpha \leftrightarrow 1-\alpha$, a symmetry which is not manifest in our notation. In our notation the replacement $2\leftrightarrow 3$ should be accompanied by a simple redefinition of the fermions.}
\begin{align}
\label{L3-full1}
\nonumber
\mathcal{L}_3 &= \frac{1}{\sqrt{2}}\sqrt{\alpha(1-\alpha)}\Big[-\alpha\big(\bar\chi_-^4\bar\chi_-^2-\bar\chi_-^1\bar\chi_-^3
+\bar\chi_+^1\bar\chi_+^3-\bar\chi_+^4\bar\chi_+^2\big)y_2 \\
& \phantom{\frac{1}{2\sqrt{2}}\sin 2\phi\Big[\quad} -i(1-\alpha)\big(\chi_-^3\bar\chi_-^4+\chi_-^2\bar\chi_-^1+\chi_+^3\bar\chi_+^4+\chi_+^2\bar\chi_+^1\big)y_3\\ \nn
& \phantom{\frac{1}{2\sqrt{2}}\sin 2\phi\Big[\quad} -2\big(\chi_-^2\bar\chi_+^3+\chi_+^2\bar\chi_-^3\big)y'_1 + 2\big(\chi_-^2\bar\chi_+^2-\chi_+^3\bar\chi_-^3\big)\dot{y}_4 \\ \nn
& \phantom{\frac{1}{2\sqrt{2}}\sin 2\phi\Big[\quad} + \big(\chi_-^3\bar\chi_+^4-\chi_-^2\bar\chi_+^1\big)(\dot{y}_3+y'_3) + \big(\chi_+^3\bar\chi_-^4-\chi_+^2\bar\chi_-^1\big)(\dot{y}_3-y'_3)\\ \nn 
& \phantom{\frac{1}{2\sqrt{2}}\sin 2\phi\Big[\quad} + i\big(\bar\chi_-^3\bar\chi_+^1+\bar\chi_-^2\bar\chi_+^4\big)(\dot{y}_2+y'_2) + i\big(\bar\chi_-^1\bar\chi_+^3+\bar\chi_-^4\bar\chi_+^2\big)(\dot{y}_2-y'_2)\Big]\\ \nn
& \phantom{\quad} -\sqrt{2}\sqrt{\alpha(1-\alpha)}\big(\alpha|y_2|^2-(1-\alpha)|y_3|^2\big)\ \dot{y}_4 + \text{h.c.}\ ,
\end{align}
where the hermitian conjugate is defined in the standard way, $\big(\chi_-\bar\chi_+\big)^\dagger = \chi_+\bar\chi_-$ etc. Here $\partial_0$ and $\partial_1$ are denoted by dots and primes respectively. Note that for the $\alpha=0$ and $\alpha=1$ cases, corresponding to $AdS_3\times S^3\times T^4$, the cubic Lagrangian vanishes. This simplifies many computations in the $T^4$ case as there are then no 3-vertices. 

In order for the light-cone gauge (\ref{eq-gauge}) to be valid beyond the cubic approximation, higher order corrections to the worldsheet metric are needed. These enter in the form
\bea 
\gamma_{ij}=\eta_{ij}+\frac{1}{g}\gamma^{1}_{ij}+\frac{1}{g^2}\gamma^{2}_{ij}+\dots
\eea 
where the subleading pieces are quadratic and quartic in fields respectively. For general $a$, the corrections are rather involved but they simplify in the uniform gauge. For example, the leading order corrections for $a=\frac{1}{2}$ are
\bea \nn
\gamma^1_{00}=1-|y_1|^2+\alpha^2|y_2|^2+(1-\alpha)^2|y_3|^2,\qquad 
\gamma^1_{11}=-1-|y_1|^2+\alpha^2|y_2|^2+(1-\alpha)^2|y_3|^2
\eea 
and $\gamma^1_{01}=\gamma^1_{10}=0$. The subleading pieces are involved but straightforwardly determined. Having fixed the worldsheet metric so that the equation of motion for $x^-$ is satisfied, one can in principle write down the Lagrangian to any order in perturbation theory. Naturally, for general $a$ and $\alpha$ the structural complexity is rather daunting. Here we will only write the purely bosonic terms in the quartic Lagrangian
which, in the $a=\frac{1}{2}$ gauge, take the form
\begin{align}
\label{eq:l4_bosonig}
\nn
\mathcal{L}_4^{B} &= \alpha(1-\alpha)\left(\alpha |y_2|^2-(1-\alpha) |y_3|^2\right)^2 - \frac{1}{2}\alpha(1-\alpha)\left(\dot{y}_4^2 + \dot{\bar{y}}_4^2 - y_4'^2 - \bar{y}_4'^2\right)\left(|y_2|^2 + |y_3|^2\right)\\ \nn 
&\phantom{\quad} - |\dot{y}_4|^2\left(|y_1|^2 - (2\alpha-1)^2(\alpha|y_2|^2 - (1-\alpha)|y_3|^2)\right) + |\dot{y}_1|^2\big(\alpha^2|y_2|^2 + (1-\alpha)^2|y_3|^2\big) \\ \nn
&\phantom{\quad} - (|\dot{y}_2|^2 + |\dot{y}_3|^2 + |y'_i|^2)\left(|y_1|^2 - \alpha^2|y_2|^2 - (1-\alpha)^2|y_3|^2\right) - \alpha|\dot{y}_2|^2|y_2|^2-(1-\alpha)|\dot{y}_3|^2|y_3|^2\\ \nn
&\phantom{\quad} + \alpha(1-\alpha)|y'_4|^2(|y_2|^2 + |y_3|^2) - |y'_1|^2|y_1|^2 + \alpha|y'_2|^2|y_2|^2 + (1-\alpha)|y'_3|^2|y_3|^2\,.\\
\end{align} 
The relevant terms for our computations which mixes bosons and fermions can be found in (\ref{Lbf}).

\section{Tree-level scattering in \3}
\label{sec:tree-level-scattering}
Following the notation of \cite{Zarembo:2009au}, the worldsheet S-matrix separates into scattering, transmission and reflection pieces,
\begin{align*}
 \mathbbm{S}&=\mathbbm{1}+\frac{i}{g}S+...  \,\,=1+\qquad\parbox[top][0.8in][c]{1in}{\fmfreuse{S-t-chan}} + \qquad \parbox[top][0.8in][c]{1in}{\fmfreuse{S-u-chan}} + \qquad\parbox[top][0.8in][c]{1.5in}{\fmfreuse{S-c}} \\ \nn 
\mathbbm{T}&=\mathbbm{1}+\frac{i}{g}T+...\,\,  =1+\qquad\parbox[top][0.8in][c]{1in}{\fmfreuse{T-t-chan}} + \qquad \parbox[top][0.8in][c]{1in}{\fmfreuse{T-s-chan}}+ \qquad\parbox[top][0.8in][c]{1.5in}{\fmfreuse{T-c}}
 \\ \nn 
\mathbbm{R}&=\frac{i}{g}R+... \qquad =\qquad \qquad\parbox[top][0.8in][c]{1in}{\fmfreuse{R-t-chan}} + \qquad \parbox[top][0.8in][c]{1in}{\fmfreuse{R-s-chan}}+ \qquad\parbox[top][0.8in][c]{1.5in}{\fmfreuse{R-c}}
\end{align*}
We will consider processes with two incoming bosons and we write the relevant S-matrix elements as
\bea 
&& S\cdot\ket{y_i(p_1) y_j(p_2)}=A^{(ij,kl)}\ket{y_k y_l}+B^{(ij,kl)}\ket{\chi_k \chi_l}, \\ \nn 
&& T\cdot\ket{y_i(p_1)\bar y_j(p_2)}=A^{(i\bar{j},k\bar l)}\ket{y_k\bar y_l}+B^{(i\bar j,k\bar l)}\ket{\chi_k \bar \chi_l}, \\ \nn 
&& R\cdot\ket{y_i(p_1)\bar y_j(p_2)}=\tilde A^{(i \bar{j},\bar kl)}\ket{\bar y_k y_l}+\tilde B^{(i\bar j,\bar kl)}\ket{ \bar \chi_k\chi_l}
\eea 
where we will take $i,j=2,3$ to be any of the two light particles with masses $\alpha$ and $(1-\alpha)$ respectively. When massless particles are involved a somewhat more general structure is allowed with $A^{i\bar j,kl}$ etc. Note that many amplitudes are trivially zero due to $U(1)$-invariance. 

It will turn out that for all the amplitudes we compute the $R$-piece will be zero and we will not write it out explicitly.

\subsubsection*{Diagonal light processes in general $a$-gauge}
In \cite{Rughoonauth:2012qd} the diagonal elements $A^{(ij)}=A^{(ij,ij)}$ for $i,j=2,3$ were determined in the uniform light-cone gauge $a=\frac{1}{2}$. We will begin by generalizing this result to a general a-gauge. Summing up the $s,t$ and $u$-channel contributions together with the contact terms we find
\bea 
\label{eq:tree-level-S}
&&A^{(22)}=1+\frac{i}{2}\frac{\alpha\big(p_1+p_2\big)^2}{\omega^{(2)}_1 p_2-\omega^{(2)}_2 p_1}+\frac{i}{2}(1-2a)\big(\omega^{(2)}_1 p_2-\omega^{(2)}_2 p_1\big), \\ \nn 
&&  A^{(2\bar2)}=1+\frac{i}{2}\frac{\alpha\big(p_1-p_2\big)^2}{\omega^{(2)}_1 p_2-\omega^{(2)}_2 p_1}+\frac{i}{2}(1-2a)\big(\omega^{(2)}_1 p_2-\omega^{(2)}_2 p_1\big) \\ \nn 
&&  A^{(23)}=A^{(2\bar3)}=1+\frac{i}{2}(1-2a)\big(\omega^{(2)}_1 p_2-\omega^{(3)}_2 p_1\big)\,.
\eea 
The $33$, $3\bar 3$, $\bar2\bar3$ and $\bar23$ elements are given by sending $\alpha\rightarrow 1-\alpha$ in an obvious way. We have used the fact that the delta functions from energy momentum conservation force\footnote{With particles of different masses, this relation can become significantly more complicated.} 
\bea 
\label{eq:momentum-convention}
\vec p_3=\vec p_1,\qquad \vec p_4=\vec p_2
\eea 
and the relativistic on-shell energies above are explicitly given by
\bea \nn 
\omega^{(1)}(p)=\sqrt{1+p^2},\quad 
\omega^{(2)}(p)=\sqrt{\alpha^2+p^2},\quad 
\omega^{(3)}(p)=\sqrt{(1-\alpha)^2+p^2},\quad 
\omega^{(4)}(p)=|p|
\eea
In the above equations we have included the external leg factors and Jacobian from the overall delta functions,
\bea  \label{eq:leg-jacobian-factor}
\frac{1}{4\big(\omega_1^{(i)}p_2-\omega_2^{(j)} p_1\big)}\,.
\eea 
We now turn to the non-diagonal elements in the scattering of two light bosons, as some of these are rather involved we will only evaluate them in the uniform gauge $a=\frac12$.

\subsubsection*{Non-diagonal processes in uniform gauge}
Here we will consider processes where the incoming particles are light-bosons but the out-going particles can be anything allowed by conservation of the $U(1)$ charges. For transition type amplitudes, where the in-state is neutral, there are typically several possible out-states. What is more, when the masses of the scattered particles are different the solution of the energy momentum constraint becomes more complicated. Since some amplitudes are rather involved already at tree-level we will restrict to $\alpha=1$ in these cases. Collecting the various contributions we have\footnote{To see that processes involving particles of different mass, for example $A^{(2\bar2,3\bar3)}$ and $A^{(2\bar2,1\bar1)}$, are zero a useful identity which follows from energy-momentum conservation is
$$
m_f^2(\omega_1\omega_2+p_1p_2)=m_i^2(\omega_3\omega_4+p_3p_4)\,,
$$
where the incoming particles have mass $m_i$ and the outgoing ones mass $m_f$.
}
\bea 
&& \ket{y_2\bar y_2}\quad \rightarrow \quad A^{(2\bar2,i\bar j)} \ket{y_i\bar y_j} + B^{(2\bar 2,i\bar j)}\ket{\chi_i \bar \chi_j}\,: \nn \\ 
&&\nn\\
&& A^{(2\bar2,1\bar 1)}=0,\qquad A^{(2\bar2,3\bar 3)}=0,\qquad A^{(2\bar2,4\bar 4)}=A^{(2\bar2,44)}=A^{(2\bar2,\bar4\bar4)}=0\,,\nn\\
&& B^{(2\bar2,1\bar 1)}\vert_{\alpha=1}=-\frac{i}{4}\frac{\big(1-p_-^2\big)\big(1-(p'_-)^2\big)}{	\sqrt{p_- p'_-}\big(p_-+p'_-\big)}\,,\qquad B^{(2\bar2,2\bar 2)}=-\frac{i}{4\alpha}\frac{\big(\alpha^2-p_-^2\big)\big(\alpha^2-(p'_-)^2\big)}{	\sqrt{p_- p'_-}\big(p_-+p'_-\big)}\,, \nn\\
&& B^{(2\bar2,3\bar 3)}=B^{(2\bar2,4\bar 4)}=B^{(2\bar2,44)}=B^{(2\bar2,\bar4\bar4)}=0\,,
\label{eq:AandBs}
\eea
where $p_-=p^1_-$ and $p'_-=p^2_-$ and the light-cone momenta satisfy $p_+=\frac{m^2}{p_-}$ since the particles are on-shell. Later we will utilize the so-called Near Flat Space, or Maldacena-Swanson, limit \cite{Maldacena:2006rv} where the right moving sector is boosted and we choose the right moving notation here to establish a simple connection with the results presented there\footnote{All expressions presented in this section are however for the BMN string, albeit in a right moving notation.}. In the mixed sector we find
\bea 
{B^{(2\bar3,23)}=-\frac{(\alpha^2-p_-^2)((1-\alpha)^2-(p'_-)^2)}{4\sqrt{p_-p'_-}\,((1-\alpha)p_-+\alpha p'_-)}\,,\qquad }
B^{(23,14)}\vert_{\alpha=1}=\frac{1}{4}\frac{p'_-\big(1-p_-^2\big)}{\sqrt{p_- p'_-}}
\label{eq:AandBs2}
\eea
Let us also note that in the purely bosonic sector there is no mixing between the light and the massless modes at tree-level, i.e.
\begin{equation}
A^{(2\bar2,44)}=A^{(2\bar2,4\bar4)}=A^{(24,24)}=A^{(24,2\bar4)}=0\qquad\mbox{etc.}
\end{equation}
This follows, for example, from (\ref{eq:AandBs}) by crossing.

\section{One-loop scattering in $AdS_3\times S^3\times T^4$}
\label{sec:one-loop-scattering}
\begin{figure}[t]
\centering
\includegraphics[scale=0.8]{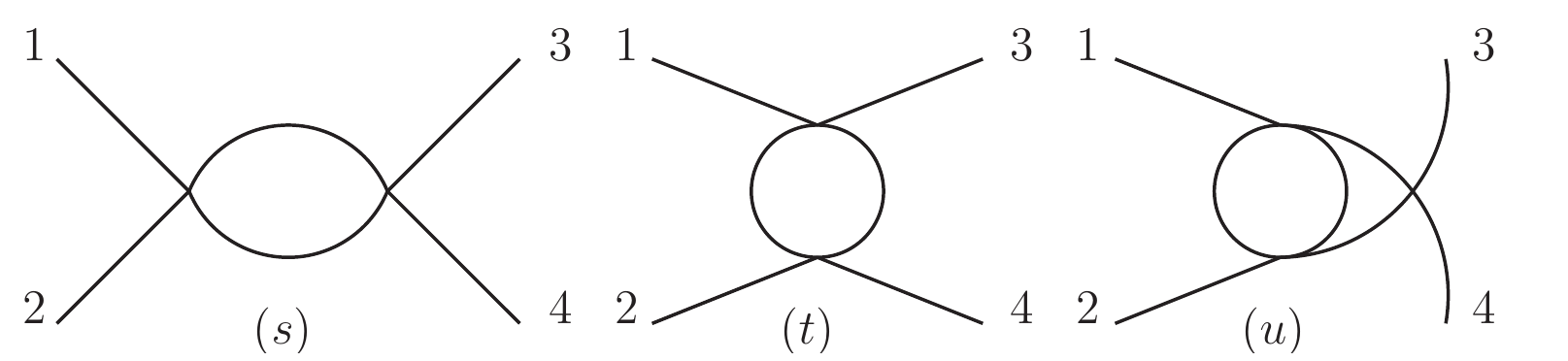}
\caption{The $s, t$ and $u$-channel diagrams. We also have six-vertex contact diagrams but these have trivial topology}
\label{fig:stu}
\end{figure}

Generally we have both cubic, quartic and sixth order vertices implying that the topology of a generic one-loop amplitude can be rather involved. In order to reduce the number of contributing topologies we will work in $AdS_3\times S^3\times T^4$, i.e. $\alpha=1$, and take the $a=\frac{1}{2}$ gauge . In this limit the third (complex) coordinate $y_3$ and $\chi^3_\pm$ become massless and the cubic Lagrangian vanishes. Equivalently we could of course take $\alpha=0$ where $y_2$ and $\chi^2_\pm$ become massless. In both cases the contributing diagrams are just of four distinct types, see figure \ref{fig:stu}. 

\subsection{One-loop amplitudes in the massless sector}
\label{sec:subsection-massless-one-loop}
In this section we will compute the $S,T$ and $R$ matrices for massless bosonic $2\rightarrow 2$ processes. For these modes the dispersion relation is simply $\epsilon=|p|$ and as it turns out the final answer will take a rather simple form. Furthermore, the relevant part of the quartic BMN Lagrangian simplifies to
\bea \label{eq:L4rel}
&& 
\mathcal{L}_4=\\ 
\nn 
&&-\frac{1}{2}\big(\partial_+y_3\partial_-\bar y_3+\partial_+y_4\partial_-y_4\big)\big(\bar\chi^4_+\chi^4_-+\bar\chi^3_-\chi^3_+\big)
-\frac{1}{2}\big(\partial_-y_3\partial_+\bar y_3+\partial_+\bar y_4\partial_-\bar y_4\big)\big(\bar\chi^4_-\chi^4_++\bar\chi^3_+\chi^3_-\big) \\ \nn 
&&
-\frac{1}{2}\big(\partial_+\bar y_3\partial_+ y_3+\partial_-\bar y_3\partial_- y_3+\partial_+\bar y_4\partial_+ y_4+\partial_-\bar y_4\partial_- y_4\big)(|y_1|^2-|y_2|^2) \\ \nn 
&& 
-\frac{1}{2}\big(\partial_- y_4\partial_+\bar y_3-\partial_+\bar y_4\partial_-\bar y_3\big)\big(\bar \chi^3_+\chi^4_--\bar \chi^3_-\chi^4_+\big)
-\frac{1}{2}\big(\partial_+ y_4\partial_- y_3-\partial_+ y_3\partial_-\bar y_4\big)\big(\bar \chi^4_+\chi^3_--\bar \chi^4_-\chi^3_+\big)
\eea 
Thus the one-loop diagrams with all massive (or massless) external particles will have either two massive modes or two massless modes propagating in the loop. 

The full sixth order BMN Lagrangian is naturally rather involved, but as it turns out we only need the purely bosonic part for massless processes,
\bea 
\mathcal{L}_6=-\frac{1}{4}\big(\partial_+ y_3\partial_+\bar y_3+\partial_+y_4\partial_+\bar y_4\big)\big(\partial_- y_3\partial_-\bar y_3+\partial_-y_4\partial_-\bar y_4\big)\Big(|y_1|^2+|y_2|^2\Big)+\dots
\eea 
Furthemore, since the particles are massless it makes sense to demand that the sign of the incoming momenta should be opposite, otherwise they would never meet. Hence we will assume $p_1>0>p_2$. Evaluating the integrals in (\ref{eq:massless33})-(\ref{eq:massless43}) using (\ref{eq:standard-integrals}) with the notation $p_\pm =p^1_\pm$ and $p'_\pm=p^2_\pm$ gives
\bea 
&& S^{(1)}_{33}= \frac{i}{8\pi}\Big[\big(p_+ p'_-\big)^2\Big]_t-\frac{i}{8\pi}\Big[\Big(-\frac{2}{\epsilon}+\gamma+\log\pi-\log p'_- p_+ \Big)\big(p_+ p'_-\big)^2\Big]_u \\ \nn 
&& +\frac{i}{8\pi}\Big[\Big(-\frac{2}{\epsilon}+\gamma+\log\pi \Big)\big(p_+ p'_-\big)^2\Big]_c
=\frac{2 i}{\pi}\big(1-\log\big[ -4 p_1 p_2\big]\big)p_1^2p_2^2, \\ \nn 
&&\nn\\
&& T^{(1)}_{3\bar 3}=\frac{i}{8\pi}\Big[\big(p_+ p'_-\big)^2\Big]_t-\frac{i}{8\pi}\Big[\Big(-\frac{2}{\epsilon}+\gamma+\log\pi-\log -p'_- p_+ \Big)\big(p_+ p'_-\big)^2\Big]_s \\ \nn 
&& +\frac{i}{8\pi}\Big[\Big(-\frac{2}{\epsilon}+\gamma+\log\pi \Big)\big(p_+ p'_-\big)^2\Big]_c=\frac{2 i}{\pi}\big(1-\log\big[ 4 p_1 p_2\big]\big)p_1^2p_2^2,\quad R^{(1)}_{3\bar 3}=0,\\ \nn 
&&\nn\\
&& S^{(1)}_{44}=\frac{2 i}{\pi}\big(1-\log\big[ 4 p_1 p_2\big]\big)p_1^2p_2^2,\quad T^{(1)}_{4\bar 4}=\frac{2 i}{\pi}\big(1-\log\big[ -4 p_1 p_2\big]\big)p_1^2p_2^2,\quad R^{(1)}_{4\bar 4}=0, \\ \nn 
&&\nn\\
&& S^{(1)}_{43}=\frac{2 i}{\pi}\big(1-\log\big[ 4 p_1 p_2\big]\big)p_1^2p_2^2,\quad T^{(1)}_{4\bar 3}=\frac{2 i}{\pi}\big(1-\log\big[ -4 p_1 p_2\big]\big)p_1^2p_2^2,\quad R^{(1)}_{4\bar 3}=0
\eea 
where the $c$ index denotes contact terms arising from six-vertex interactions. We have ignored the external leg factors and Jacobian from the overall delta functions. Furthermore, note that in the NFS limit, see (\ref{eq:NFS-limit-appendix}) for details, the amplitudes are zero since $p_+ p'_-\sim \mathcal{O}(1)$.

Since $p_1 p_2<0$ (or equivalently $p_-< p'_-$) we see that some of the processes develop a real part which breaks the $SO(4)$ invariance. This should be due to the fact that we are working in type IIA where there is flux along one of the $T^4$-directions. T-dualizing the $x^9$-coordinate one should recover the (manifest) $SO(4)$-invariance of the type IIB case. We will leave further investigation of the massless processes for an upcoming paper. Especially interesting is of course to understand the apparent mismatch with the conjectured set of Bethe equations for massless fields in \cite{Sax:2012jv}.

\subsection{Massive one-loop scattering in NFS limit}
We now wish to look at the scattering of the massive bosonic modes in $AdS_3\times S^3\times T^4$ ($\alpha=1$). However, at the one loop level one encounters problems with the BMN string since the amplitudes (naively) fail to be UV finite, a problem which also appears for the more symmetric $AdS_5 \times S^5$ string\footnote{We thank T. McLoughlin for making us aware of this (unpublished) result. The reason this problem did not appear already in section \ref{sec:subsection-massless-one-loop} is probably because there only the purely bosonic piece of the sixth order Lagrangian contributed.} This is most likely related to some subtlety with the light-cone gauge fixing we employ. Luckily, the problem can be circumvented by utilizing the Near Flat Space, or Maldacena-Swanson, limit \cite{Maldacena:2006rv}. The details of the limit can be found in appendix A, see especially (\ref{eq:NFS-limit-appendix}). The limit basically boils down to a boost of the right-moving (worldsheet) sector and all divergent loop integrals are shifted to subleading orders in the $1/g$ expansion. Furthermore, simple power counting demonstrates that only quartic vertex diagrams contribute, so the topologies of Feynman diagrams are completely summarized in figure \ref{fig:stu}. 

In $\alpha=1$, the massive modes are the $1$ and $2$ particles. Using the expressions in (\ref{eq:appendix-massive-integrals}) we find \footnote{We have taken out a factor of $1/(16\pi)$ and $\big(p'_-p_-\big)$ from the $s,t$ and $u$-channel contributions in the first lines of $S^{(1)}_{ii}$ and $T^{(1)}_{ii}$.}
\bea \label{eq:NFS-one-loop}
&& S^{(1)}_{ii}= \frac{1}{32\pi}\frac{\big(p_- p'_-\big)^2}{(p'_-)^2-p_-^2}\times\\ \nn 
&& \Big(\Big[i\big(p_-+p'_-\big)^2\Big]_t
+\Big[\frac{i(p'_-+p_-)\big((p'_-)^2+p_-^2\big)}{p'_--p_-}\log\frac{p'_-}{p_-}\Big]_u
-\Big[\frac{\big(p'_-+p_-\big)^3}{p'_--p_-}\big(\pi+i\log\frac{p'_-}{p_-}\big)\Big]_s\Big)
\\ \nn 
&& =\frac{1}{32\pi}\frac{(p_-p'_-)^2}{(p'_-)^2-p_-^2}\Big[i\big(p_-+p'_-\big)^2-\frac{2ip_-p'_-(p_-+p'_-)\log\frac{p'_-}{p_-}}{p'_--p_-}-\frac{\pi\big(p_-+p'_-\big)^3}{p'_--p_-}\Big], \\ \nn 
&& T^{(1)}_{ii}= \frac{1}{32\pi}\frac{\big(p_- p'_-\big)^2}{(p'_-)^2-p_-^2}\times \\ \nn 
&& \Big(-\Big[i\big(p'_--p_-\big)^2\Big]_t
+\Big[\frac{i\big(p'_--p_-\big)^3}{p'_-+p_-}\log\frac{p'_-}{p_-}\Big]_u
-\Big[\frac{(p'_--p_-)\big((p'_-)^2+p_-^2\big)}{p'_-+p_-}\big(\pi+i\log\frac{p'_-}{p_-}\big)\Big]_s\Big)
\\ \nn 
&&= 
-\frac{1}{32\pi}\frac{(p_-p'_-)^2}{(p'_-)^2-p_-^2}\Big[i\big(p_--p'_-\big)^2+\frac{2ip_-p'_-(p'_--p_-)\log\frac{p'_-}{p_-}}{p'_-+p_-}+\frac{\pi(p'_--p_-)\big(p_-^2+(p'_-)^2\big)}{p_-+p'_-}\Big], \\ \nn 
&& S^{(1)}_{ij}=-\frac{1}{32\pi}\frac{(p_-p'_-)^2}{(p'_-)^2-p_-^2}\Big[i\big(p_--p'_-\big)^2+\frac{2ip_-p'_-(p'_--p_-)\log\frac{p'_-}{p_-}}{p'_-+p_-}+\pi\big((p'_-)^2-p_-^2\big)\Big],
\quad i\neq j \\ \nn 
&& 
T^{(1)}_{ij}=\frac{1}{32\pi}\frac{(p_-p'_-)^2}{(p'_-)^2-p_-^2}\Big[i\big(p_-+p'_-\big)^2-\frac{2ip_-p'_-(p'_-+p_-)\log\frac{p'_-}{p_-}}{p'_--p_-}-\frac{\pi(p_-+p'_-)\big(p_-^2+(p'_-)^2\big)}{p'_--p_-}\Big], 
\quad i\neq j \\ \nn 
\eea 
where we, in contrast to the massless case, also included the external leg factors and the Jacobian from the delta function, which in right moving components equals
\bea \nn 
\frac{1}{2}\frac{p_- p'_-}{(p'_-)^2-p_-^2}\,.
\eea 
While we have not written out the explicit contributions, one can also show that the $s,t$ and $u$-channel contributions in the reflection piece, $R_{ij}^{(1)}$, cancel among themselves. Hence, at least for $\alpha=1$ the one-loop S-matrix seems to be reflectionless, a fact so far only seen at tree-level in \cite{Rughoonauth:2012qd}.

%%%%%%%%%%%%%%%%%%%%%%%%%%%%%%%%%%%%%%%%%%%%%%%%%%%%%%%%%%%%%%%%%%%%%%%%%%%%%%%%%%%%%%%%%%%%%%%%%%

\section{Comparison to proposed exact S-matrix}
\label{sec:exact-s-matris}
Here we will compare our results from worldsheet calculations to the proposals for the exact S-matrix \cite{Borsato:2012ss, Ahn:2012hw}. Since we are looking only at the scattering of bosons the relevant terms in the S-matrix are
\begin{equation}
S|y_iy_j\rangle=A^{(ij)}|y_iy_j\rangle+B^{(ij)}|\chi_i\chi_j\rangle\,.
\end{equation}
For the S-matrix proposed in \cite{Borsato:2012ss} these take the form
\begin{eqnarray}
A_{12}^{(22)}&=&(S_{12}^{(22)})^{-1}\frac{x_1^--x_2^+}{x_1^+-x_2^-}\nonumber\\
A_{12}^{(2\bar2)}&=&(S_{12}^{(2\bar2)})^{-1}
\frac{1-\frac{1}{x_1^+x_2^-}}{\sqrt{\left(1-\frac{1}{x_1^+x_2^+}\right)\left(1-\frac{1}{x_1^-x_2^-}\right)}}
%=1-i/4(p_1-p_2)-i/4\alpha(p_1-p_2)^2/(e_1p_2-e_2p_1)
\label{eq-su11-elements}
\\
A_{12}^{(23)}&=&(S_{12}^{(23)})^{-1}\nonumber\\
A_{12}^{(2\bar3)}&=&(S_{12}^{(2\bar3)})^{-1}\frac{1-\frac{1}{x_1^+y_2^-}}{\sqrt{\left(1-\frac{1}{x_1^+y_2^+}\right)\left(1-\frac{1}{x_1^-y_2^-}\right)}}
\nonumber
\end{eqnarray}
together with
\begin{eqnarray}
B_{12}^{(2\bar2)}&=&-(S_{12}^{(2\bar2)})^{-1}
\frac{\sqrt{-(x_1^--x_1^+)(x_2^--x_2^+)}}{x_1^-x_2^-}
\frac{1}{\sqrt{\left(1-\frac{1}{x_1^+x_2^+}\right)\left(1-\frac{1}{x_1^-x_2^-}\right)}}
%((\alpha-e_1)(\alpha-e_2))^{1/2}
%1/2-\alpha/2\frac{p_1-p_2}{e_1p_2-e_2p_1}
\nonumber\\
B_{12}^{(2\bar3)}&=&-(S_{12}^{(2\bar3)})^{-1}
\frac{\sqrt{-(x_1^--x_1^+)(y_2^--y_2^+)}}{x_1^-y_2^-}
\frac{1}{\sqrt{\left(1-\frac{1}{x_1^+y_2^+}\right)\left(1-\frac{1}{x_1^-y_2^-}\right)}}
%((\alpha-e_1)(1-\alpha-e_2))^{1/2}
%1/2-1/2\frac{(1-\alpha)p_1-\alpha p_2}{e_1p_2-e_2p_1}
\label{eq:Bs}
\end{eqnarray}
where particle 2 and 3 have mass $\alpha$ and $1-\alpha$ respectively. The other elements are obtained by $2\leftrightarrow3$ and $x\leftrightarrow y$. The phases appearing in these expressions are given by
\begin{eqnarray}
&& S_{12}^{(22)}=S_{12}^{(2\bar2)}\left(\frac{1-\frac{1}{x_1^+x_2^-}}{1-\frac{1}{x_1^-x_2^+}}\right)^{1/2}
=
e^{-i\frac{\tilde a}{\alpha h}(\omega_1p_2-\omega_2p_1)}
\frac{x_1^-}{x_1^+}\frac{x_2^+}{x_2^-}
\frac{1-\frac{1}{x_1^+x_2^-}}{1-\frac{1}{x_1^-x_2^+}}
\sigma^2(x_1,x_2)
%%%=
%(1+i/2(e_2p_1-e_1p_2)/\alpha)
%(1-i(p_1-p_2))
%(1-i/2(p_1-p_2)-i/2\alpha(p_1-p_2)^2/(e_1p_2-e_2p_1))
%(1+i(p_1-p_2)-i/2\alpha(p_1-p_2)^2/(e_2p_1-e_1p_2)-i/2(e_2p_1-e_1p_2)/\alpha)
%=(1-i/2(p_1-p_2))
\nn \\ 
&& S_{12}^{(23)}=S_{12}^{(2\bar3)}\left(\frac{1-\frac{1}{x_1^+y_2^-}}{1-\frac{1}{x_1^-y_2^+}}\right)^{-1/2}
=
e^{-\frac{i}{h}(\frac{\tilde a}{\alpha}\omega_1p_2-\frac{\tilde b}{1-\alpha}\omega_2p_1)}
\left(\frac{x_1^-}{x_1^+}\frac{y_2^+}{y_2^-}\right)^{1/2}
\frac{1-\frac{1}{x_1^+y_2^-}}{1-\frac{1}{x_1^-y_2^+}}
\sigma^2(x_1,y_2)\,.
%1-\frac{i}{h}(\frac{\tilde a}{\alpha}\omega_1p_2-\frac{\tilde b}{1-\alpha}\omega_2p_1)
%1-i(p_1-p_2)/2
%1-i/2(p_1-p_2)-i/2((1-a)p_1-ap_2)\frac{(p_1-p_2)}{e_1p_2-e_2p_1}
%1-i(p_j-p_k)\Big[1+\frac{(1-\alpha)p_k-\alpha p_j}{p_j\omega_k-p_k\omega_j}\Big]-\frac{i}{4\alpha(1-\alpha)}\Big[\frac{((1-\alpha)p_k-\alpha p_j)^2}{p_j\omega_k-p_k\omega_j}+p_k\omega_j-p_j\omega_k\Big]
%=
%1+\frac{i}{2}(e_1p_2/a-e_2p_1/(1-a))-i(\frac{\tilde a}{\alpha}e_1p_2-\frac{\tilde b}{1-\alpha}e_2p_1)
\nn \\
\label{eq:S1222}
\end{eqnarray}
Note that we have chosen the free parameters $\gamma$ and $\Gamma$ appearing in \cite{Borsato:2012ss} to be zero so as to have the standard factor of $\sigma^2$ in $S^{(22)}$ that one expects to have \cite{Babichenko:2009dk}. In \cite{Ahn:2012hw} $S^{(23)}$ was taken to be simply $1$, corresponding to $\Gamma=-1/2$ in \cite{Borsato:2012ss}, but this does not seem to give the right answer in a general $a$-gauge as we will discuss below. We have also translated to the string basis by multiplying with $\left(\frac{x_1^-}{x_1^+}\frac{x_2^+}{x_2^-}\right)^{1/2}$. For this choice of parameters the purely bosonic S-matrix also agrees with that of \cite{Ahn:2012hw}, modulo the issue with $S^{(23)}$, provided that one makes the replacement,
\begin{equation}
\frac{1-\frac{1}{x_1^+x_2^-}}{\sqrt{\left(1-\frac{1}{x_1^+x_2^+}\right)\left(1-\frac{1}{x_1^-x_2^-}\right)}}
\rightarrow
\sqrt{\frac{1-\frac{1}{x_1^+x_2^-}}{1-\frac{1}{x_1^-x_2^+}}}
\end{equation}
in $A_{12}^{(2\bar2)}$ and similarly in $A_{12}^{(2\bar3)}$. {For the S-matrix of \cite{Ahn:2012hw} the $B$ coefficients vanish for $\alpha\neq0,1$. This clearly disagrees with the results of our worldsheet computations since we find $B\neq0$.}

Note that we have included in (\ref{eq:S1222}) gauge-dependent phases \cite{Arutyunov:2009ga} involving parameters $\tilde a$ and $\tilde b$, which will be related to the worldsheet $a$-gauge parameter below. These gauge-dependent phases were not included in the analysis of \cite{Borsato:2012ss} but we seem to find that they are needed in order to match our worldsheet results in a general $a$-gauge.

The phase $\sigma$ appearing in (\ref{eq:S1222}) is required to reduce to the (modified) AFS phase, discussed below, at lowest order.

\subsection{Tree-level}
Here we will expand the S-matrix elements in (\ref{eq-su11-elements}) and (\ref{eq:Bs}) to lowest order in $h(\lambda)=\sqrt{\frac{\lambda}{2}}+\mathcal O(\lambda^0)$ and compare them to the worldsheet tree-level results found earlier. Using the fact that
\begin{equation}
\label{xpm}
x^\pm=\frac{\alpha+\sqrt{\alpha^2+4h(\lambda)^2\sin^2\frac{p}{\sqrt{2\lambda}}}}{2h(\lambda)\sin\frac{p}{\sqrt{2\lambda}}}e^{\pm i\frac{p}{\sqrt{2\lambda}}}
=\frac{\alpha+\omega_p}{p}\left(1\pm i\frac{p}{\sqrt{2\lambda}}\right)+\mathcal O(\frac{1}{\lambda})\,,
\end{equation}
and similarly for $y^\pm$ with $\alpha\rightarrow1-\alpha$, we get
\begin{eqnarray}
A_{12}^{(22)}&=&1-\frac{i}{2h}\frac{\alpha(p_1+p_2)^2}{\omega_1p_2-\omega_2p_1}
-\frac{i}{2h\alpha}(1-2\tilde a)(\omega_1p_2-\omega_2p_1)
\,,\nn\\
A_{12}^{(2\bar2)}&=&1-\frac{i}{2h}\frac{\alpha(p_1-p_2)^2}{\omega_1p_2-\omega_2p_1}
-\frac{i}{2h\alpha}(1-2\tilde a)(\omega_1p_2-\omega_2p_1)
\,,\nn\\
A_{12}^{(23)}&=&A_{12}^{(2\bar3)}=
1-i\frac{1-2\tilde a}{2h\alpha}\omega_1p_2+i\frac{1-2\tilde b}{2h(1-\alpha)}\omega_2p_1
\,.
\end{eqnarray}
This agrees completely with the tree-level S-matrix computed in the near BMN-limit (\ref{eq:tree-level-S}) provided that $\tilde a$ and $\tilde b$ are related to the $a$-gauge parameter as 
\begin{equation}
\tilde a=\frac12(1-\alpha(1-2a))\,,\qquad\tilde b=\frac12(1-(1-\alpha)(1-2a))\,.
\end{equation}
Unless $\alpha=\frac12$ these cannot both be set to zero by a choice of $a$.

Note that for the S-matrix proposed in \cite{Ahn:2012hw} $S^{(23)}=1$ which would give $A^{(23)}=1$, disagreeing with the worldsheet calculation except when $a=\frac12$.  Modulo this issue and the gauge-dependent phases we see that the S-matrices of \cite{Ahn:2012hw} and \cite{Borsato:2012ss} match the worldsheet calculation in the bosonic sector at tree-level.

For the scattering of bosons into fermions we get for the S-matrix of \cite{Borsato:2012ss} by expanding (\ref{eq:Bs}) (setting $\tilde a=\frac12$)
\begin{eqnarray}
B_{12}^{(2\bar2)}&=&
-\frac{1}{2h}\sqrt{(\alpha-\omega_1)(\alpha-\omega_2)}\left(1-\alpha\frac{p_1-p_2}{\omega_1p_2-\omega_2p_1}\right)
=
-\frac{1}{4\alpha h}\frac{(\alpha^2-p_-^2)(\alpha^2-(p_-')^2)}{\sqrt{p_-p_-'}\,(p_-+p_-')}
%%NFS:
%-\alpha/4(p_-p'_-)^{3/2}\frac{p_--p'_-}{p_-^2-(p'_-)^2})
\nonumber\\
B_{12}^{(2\bar3)}&=&
-\frac{1}{2h}\sqrt{(\alpha-\omega_1)(1-\alpha-\omega_2)}\left(1-\frac{(1-\alpha)p_1-\alpha p_2}{\omega_1p_2-\omega_2p_1}\right)
=
-\frac{1}{4h}\frac{(\alpha^2-p_-^2)((1-\alpha)^2-(p_-')^2)}{\sqrt{p_-p_-'}((1-\alpha)p_-+\alpha p_-')}\,.
\nn\\
\end{eqnarray}
{These expressions match exactly (apart from a factor of $i$ which is due to our conventions for the fermions) those coming from the worldsheet calculation in (\ref{eq:AandBs}) and (\ref{eq:AandBs2}).

In conclusion we find complete agreement at tree-level between the worldsheet computations and the results coming from the proposed exact S-matrix of \cite{Borsato:2012ss} with gauge-dependent phases added. For the amplitudes involving fermions we disagree with the S-matrix of \cite{Ahn:2012hw}.
}

\subsection{One-loop comparison}
To lowest order the phase $\sigma$ appearing in the S-matrix is the AFS phase \cite{Arutyunov:2004vx}. Its standard all-order extension, the BES phase, does not seem to give the right answer at the one-loop order however \cite{Beisert:2006ez,Abbott:2012dd}. Instead a proposal for the one-loop correction to the phase was presented {first in \cite{David:2010yg} and then improved in \cite{Beccaria:2012kb}}. Here we will expand this phase in $h(\lambda)$ in order to compare to the one-loop NFS computation on the worldsheet. Since this computation was done in $AdS_3\times S^3\times T^4$ only we will take $\alpha=1$ here as well. The full phase has an expansion\footnote{The sum starts from $r=1$ instead of $r=2$ to account for the form of the one-loop phase of \cite{Beccaria:2012kb}.}
\bea
\label{phase-alt}
\sigma(k,j)=\exp^{i\theta_{kj}},\quad \theta_{kj}=\frac{1}{2m}\sum_{r=1}^\infty\sum_{s=r+1,r+s\,\,{\rm odd}}^\infty c_{r,s}\Big[q_r(x_k)q_s(x_j)-q_r(x_j)q_s(x_k)\Big]
\eea 
where $m=(\alpha,1-\alpha)$ for particle $2$ and $3$ respectively (in the mixed sector the phase takes a slightly different form) and\footnote{The factor of $\frac{1}{2\pi}$ in front of the one-loop term follows from our definition of the coupling.}
\bea \nn 
q_n=\frac{i}{n-1}\left(\frac{1}{(x^+)^{n-1}}-\frac{1}{(x^-)^{n-1}}\right),\qquad c_{r,s}=h(\lambda)\delta_{r+1,s}+\frac{1}{2\pi}c^{(1)}_{r,s}+\dots
\eea 
Expanding the magnon charge
\bea 
q_{n}(x_k)=\frac{p_k}{h}\Big(\frac{\omega_k-\alpha}{p_k}\Big)^{n-1}+\mathcal{O}(h^{-3})
\eea 
gives to leading order the AFS phase 
\bea 
\theta^{AFS}(x_k,x_j)&=&
\frac{h}{2\alpha}\sum_{r=2}^\infty
[
q_r(x_k)q_{r+1}(x_j)
-q_r(x_j)q_{r+1}(x_k)
]
\nn\\
&=&
-\frac{1}{4h}\left[
2(p_j-p_k)
+\alpha\frac{(p_j-p_k)^2}{\omega_jp_k-\omega_kp_j}
+\frac{1}{\alpha}(\omega_jp_k-\omega_kp_j)
\right]
+\mathcal O(h^{-3})
%or:
%\frac{\big(-1+\omega_k\big)p_j^2-\big(-1+\omega_j\big)p_k^2}{2h\big(p_k+p_j\big)}
\eea
For the mixed $(23)$-sector we use instead the form of the phase given in \cite{Borsato:2012ss}
\bea 
\theta^{AFS}(x_k,y_j)&=&-i \log\left[\left(\frac{1-\frac{1}{x^-_k y^+_j}}{1-\frac{1}{x^+_k y^-_j}}\right)\left(\frac{1-\frac{1}{x^+_k y^-_j}}{1-\frac{1}{x^+_k y^+_j}}\frac{1-\frac{1}{x^-_k y^+_j}}{1-\frac{1}{x^-_k y^-_j}}\right)^{\frac{i h}{4\alpha(1-\alpha)}\big(x^+_k+\frac{1}{x^+_k}-y^-_j-\frac{1}{y^-_j}\big)}\right]\nn\\ 
&=& 
-\frac{1}{2h}(p_j-p_k)\Big[1+\frac{(1-\alpha)p_k-\alpha p_j}{p_j\omega_k-p_k\omega_j}\Big]
\nn\\
&&{}
-\frac{1}{8h\alpha(1-\alpha)}\Big[\frac{((1-\alpha)p_k-\alpha p_j)^2}{p_j\omega_k-p_k\omega_j}+p_k\omega_j-p_j\omega_k\Big]
+\mathcal{O}(h^{-3})\,.
%\frac{i}{4h}\Big[2\big(p_k-p_j\big)+\frac{(1-2\alpha)p_k p_j}{(1-\alpha)p_k+\alpha \,p_j}\big[\frac{\omega^{(3)}_j}{1-\alpha}+\frac{\omega^{(2)}_k}{\alpha}\big]-\frac{2}{(1-\alpha)p_k+\alpha\, p_j}\big(\omega^{(3)}_j p_k^2-\omega^{(2)}_k p^2_j\big)\Big]+\mathcal{O}(h^{-3})
\eea 
Since the AFS phase is exact to order $\mathcal{O}(h^{-3})$ the one-loop contribution comes only from $c^{(1)}_{r,s}$ together with self interactions of the tree-level contribution. In the NFS limit we find for the phase defined in (\ref{eq:S1222}) that only the one-loop piece of $\sigma$ involving $c^{(1)}_{r,s}$ contributes
\begin{equation}
S_{12}^{(22)}\rightarrow(\sigma_{12}^{(1)})^2\,.
\end{equation}
%\bea 
%\Big(\frac{1-\frac{1}{x^+_p x^-_{p'}}}{1-\frac{1}{x^-_{p}x^+_{p'}}}\Big)^{1+2\gamma}=\sigma_{AFS}^{2+4\gamma}(p,p')=\dots + -\frac{(1+2\gamma)\big(p_- %p'_-\big)^2\big(p_--p'_-\big)^2}{128\big(p_-+p'_-\big)^2}\,,
%\eea 
Since this gives only an imaginary contribution the real part of the one-loop amplitude $A^{(22)}$ should therefore, according to \cite{Borsato:2012ss,Borsato:2012ud} eq. (\ref{eq-su11-elements}), be given only by
\begin{equation}
\frac{x_1^--x_2^+}{x_1^+-x_2^-}
=\dots-\frac{\big(p_- p'_-\big)^2\big(p_-+p'_-\big)^2}{32\big(p_--p'_-\big)^2}+\mathcal{O}(h^{-4})\,,
\end{equation}
where the ellipsis denote tree-level terms. This is in complete agreement with the real part found in (\ref{eq:NFS-one-loop}).

Similarly we find that the real part of $A^{(2\bar2)}$ in the NFS limit comes from
\bea \label{eq:sax-Tmatrix-el}
\Big(\frac{1-\frac{1}{x^+_1 x^-_2}}{1-\frac{1}{x^-_1 x^+_2}}\Big)^{1/2}
\frac{1-\frac{1}{x_1^+x_2^-}}{\sqrt{\left(1-\frac{1}{x_1^+x_2^+}\right)\left(1-\frac{1}{x_1^-x_2^-}\right)}}
=\ldots -\frac{(p_- p'_-)^2\big(p_-^2-p_- p'_-+(p'_-)^2\big)}{32\big(p_-+p'_-\big)^2}
\eea 
for the S-matrix of \cite{Borsato:2012ss} and
\begin{equation}
\frac{1-\frac{1}{x^+_1 x^-_2}}{1-\frac{1}{x^-_1 x^+_2}}
=
\ldots-\frac{(p_-p'_-)^2(p_--p'_-)^2}{32(p_-+p'_-)^2}
\end{equation}
for the S-matrix of \cite{Ahn:2012hw}, neither of which agree with the real part of $T_{ii}^{(1)}$ in (\ref{eq:NFS-one-loop}). The reason for this disagreement will be discussed further in the next section by applying the optical theorem.

Let us now look at the imaginary part of the amplitudes. As mentioned these can only receive contributions from the one-loop phase. In \cite{Beccaria:2012kb} the form of the one-loop contribution to the phase was suggested to be
\bea \label{eq:BL-MMT}
c_{r,s}^{(1)}=2\frac{s-r}{s+r-2},\qquad \bar c_{r,s}^{(1)}=-2\frac{s+r-2}{s-r}\,,
\eea 
where bar indicates T-type scatterings. Switching from the $k$ and $j$ labels to $p$ and $p'$ a somewhat lengthy calculation gives the NFS expansions\footnote{The first of these agrees with the phase obtained in \cite{David:2010yg}.}
\bea 
i \theta(p,p')^{\rm 1-loop}&=&
\frac{1}{2\pi}\sum \frac{s-r}{r+s-2}\Big[q_r(x)q_s(x')-q_r(x')q_s(x)\Big] \\ \nn 
&=&\frac{i}{64\pi}\frac{\big(p_-p'_-\big)^2}{(p'_-)^2-p_-^2}\Big(\frac{2p_- p'_-\big(p_-+p'_-\big)\log \frac{p'_-}{p_-}}{p'_--p_-}-\big(p_-+p'_-\big)^2\Big)+\mathcal{O}(h^{-4}) \\ \nn 
&&\\ \nn 
i\bar\theta(p,p')^{\rm 1-loop}&=&
-\frac{1}{2\pi}\sum \frac{r+s-2}{s-r}\Big[q_r(x)q_s(x')-q_r(x')q_s(x)\Big] \\ \nn 
&=&\frac{i}{64\pi}\frac{\big(p_-p'_-\big)^2}{(p'_-)^2-p_-^2}\Big(\frac{2p_- p'_-\big(p'_--p_-\big)\log \frac{p'_-}{p_-}}{p'_-+p_-}+\big(p_+-p'_-\big)^2\Big)+\mathcal{O}(h^{-4})\,.
\eea 
The imaginary terms in the amplitude are $(-2)$ times these expressions since the phase enters as $\sigma^{-2}=e^{-2i\theta}$ in (\ref{eq-su11-elements}). We see that these expressions nicely match the imaginary parts of $S^{(1)}_{ii}$ and $T^{(1)}_{ii}$ in (\ref{eq:NFS-one-loop}).

Thus we find that the expressions coming from the S-matrix proposed in \cite{Borsato:2012ss} for $22$-scattering at $\alpha=1$ and in the NFS limit match with those coming from the worldsheet calculation in (\ref{eq:NFS-one-loop}), except for the real part of $A^{(2\bar2)}$. Furthermore, to get the matching of the imaginary parts we needed to use the {one-loop phases proposed in \cite{Beccaria:2012kb} instead} of the standard Hernandez-Lopez phase \cite{Hernandez:2006tk}. We will now analyze further the reason for the discrepancy in the real part of $A^{(2\bar2)}$.

\subsection{Optical theorem}
Since the disagreement we found in the last section with the proposed S-matrix of \cite{Borsato:2012ss} is in the real part of the S-matrix element (imaginary part of the amplitude) we can use the optical theorem to trace it back to the tree-level. Taking $\alpha=1$ and the NFS limit in the tree-level results in (\ref{eq:tree-level-S}) and (\ref{eq:AandBs}) we find
\bea 
&& \ket{y_2\bar y_2}\quad \rightarrow \quad (1+A'^{(2\bar2)})\ket{y_2\bar y_2} + B^{(2\bar 2,1\bar1)}\ket{\chi_1 \bar \chi_1}+ B^{(2\bar 2,2\bar2)}\ket{\chi_2 \bar \chi_2}\,: \nn \\ \\
&&  A'^{(2\bar2)}=-\frac{i}{4}\frac{p_-p_-'(p_--p_-')}{p_-+p_-'} 
\qquad B^{(2\bar2,1\bar 1)}=B^{(2\bar2,2\bar 2)}=-\frac{i}{4}\frac{(p_-p'_-)^{3/2}}{p_-+p'_-}\,.\nn
\eea
The optical theorem then tells us that the real part of the S-matrix element for $2\bar2$-scattering should be given by
\begin{equation}
|A'^{(2\bar2)}|^2+|B^{(2\bar2,1\bar 1)}|^2+|B^{(2\bar2,2\bar 2)}|^2
=
\frac{1}{16}\frac{(p_-p_-')^2(p_-^2+(p_-')^2)}{(p_-+p_-')^2}
\end{equation}
in agreement with what we found for $T_{22}$ in (\ref{eq:NFS-one-loop}) (the relative factor of ($-2$) is due to our conventions). If we would omit the amplitude for $y_2\bar y_2\rightarrow\chi_1\bar\chi_1$ we get instead the result of \cite{Borsato:2012ss} in (\ref{eq:sax-Tmatrix-el}).

The process $y_2\bar y_2\rightarrow\chi_1\bar\chi_1$ is not accounted for in the S-matrix of \cite{Borsato:2012ss} because for $\alpha\neq0,1$ the fermion $\chi_1$ is heavy and therefore treated as a composite object \cite{Zarembo:2009au,Sundin:2012gc}. Our calculations show however that there is a non-zero amplitude for the process $y_2\bar y_2\rightarrow\chi_1\bar\chi_1$ at tree-level and this affects other one-loop S-matrix elements through the optical theorem. It is clear then that the proposed S-matrix of \cite{Borsato:2012ss} should somehow be modified in order to take this fact into account, at least in the case $\alpha=0,1$. {Indeed at $\alpha=0,1$ the $y_1$ and $\chi_1$ excitations are no longer 'heavy' but become fundamental from the Bethe ansatz perspective so one expects the exact S-matrix to include also these processes. This should resolve the one-loop discrepancy at $\alpha=1$, but from the worldsheet perspective one expects the S-matrix to change continuously with $\alpha$ so it is hard to believe that the discrepancy would not persist for $\alpha\neq0,1$.}

\section{Discussion and outlook}
We have computed tree-level worldsheet S-matrix elements for the string in $AdS_3\times S^3\times S^3\times S^1$ and certain one-loop elements in $AdS_3\times S^3\times T^4$. {The tree-level results are in agreement with those coming from the exact S-matrix proposed in \cite{Borsato:2012ss} provided one includes appropriate gauge-dependent phases. The tree-level amplitudes involving fermions disagree with those predicted from the S-matrix of \cite{Ahn:2012hw}.}

At the one-loop level we find agreement for the imaginary part of the S-matrix elements we computed provided one uses the {one-loop phases of \cite{Beccaria:2012kb} (one of which appeared in \cite{David:2010yg}) which differ} from the standard Hernandez-Lopez one \cite{Hernandez:2006tk}. This gives a nice check of their results which were derived by different methods. For the real part of the S-matrix element we seem to find a disagreement with the proposed exact S-matrices however. We have traced this issue to the fact that at tree-level there is a non-zero amplitude for two light scalars going to two heavy fermions which is not visible in the exact S-matrix, since for general $\alpha$ the heavy modes are thought of as composite. This tree-level amplitude is related to the mismatch in the real part of the S-matrix element at one-loop level through the optical theorem. This mismatch could be due to a subtlety in taking the $\alpha\rightarrow0,1$ limit in the S-matrix of \cite{Borsato:2012ss}. Indeed one expects that at $\alpha=0,1$ the modes that are heavy for $0<\alpha<1$ should no longer be composites but now appear as fundamental excitations {and the S-matrix should take this into account. It would be very interesting to see whether there is also a one-loop mismatch for $\alpha\neq0,1$ but this would require considerably more complicated calculations. From the worldsheet point of view one would expect this to be the case since the limit $\alpha\rightarrow1$ is completely smooth, at least at the level of the Lagrangian.}

We have also computed several S-matrix elements involving the massless fields. It would also be interesting to understand better how these are accounted for by the Bethe ansatz. We hope to return to this question in the near future.

Finally it would be interesting to carry out similar calculations in the less supersymmetric, and much less studied, case of $AdS_2/CFT_1$ \cite{Sorokin:2011rr,Cagnazzo:2011at,Murugan:2012mf}.

\section*{Acknowledgments}
It's a pleasure to thank M. Abbott, O. Ohlsson Sax, A. Rej, B. Stefanski and K. Zarembo for illuminating discussions. We especially want to thank B. Stefanski and K. Zarembo for participation at an early stage in the project. PS would like to thank Nordita for hospitality during the course of this work. PS is supported by a post doctoral grant from the Claude Leon Foundation. The research of LW is supported in part by NSF grant PHY-0906222.

\begin{appendix}

\section{Details of the near flat space limit}
The near flat space (NFS) limit or Maldacena-Swanson limit \cite{Maldacena:2006rv} can be considered as a worldsheet Lorentz / boost transformation together with a resummation. Originally the BMN Lagrangian has the form
\bea \label{eq:BMN-appendix}
\mathcal{L}_{BMN}=\mathcal{L}_2+\frac{1}{\sqrt{g}}\mathcal{L}_3+\frac{1}{g}\mathcal{L}_4+\dots
\eea 
where the subscript denotes number of transverse fields in each term. Performing the following worldsheet transformation\footnote{In order to make contact with the scalings of \cite{Maldacena:2006rv} one should remember that the BMN has already scaled the fermions with a factor of $g^{-1/2}$.}
\bea \label{eq:NFS-limit-appendix}
p_\pm \rightarrow g^{\mp \frac{1}{2}} p_\pm,\qquad \partial_\pm \rightarrow g^{\mp \frac{1}{2}} \partial_\pm,\qquad \chi_\pm^i\rightarrow g^{\mp \frac{1}{4}} \chi^i_\pm
\eea 
in the BMN Lagrangian and taking the $g\rightarrow \infty$ limit gives a theory without coupling dependence but still exhibiting non-trivial interactions
\bea 
\mathcal{L}_{NFS}=\mathcal{L}_2+\mathcal{L}_3+\mathcal{L}_4+\mathcal{O}(g^{-1/4})
\eea 
where the higher order terms $\mathcal{L}_3$ and $\mathcal{L}_4$ are considerably simpler than the corresponding ones in (\ref{eq:BMN-appendix}). Furthermore, all vertices beyond quartic order are suppressed in the NFS limit, simplifying it further compared to its BMN counterpart. 

\section{Integral conventions}
The integrals we encounter are standard ones
\bea \label{eq:standard-integrals}
&& B_{\mu_1\dots \mu_p}\big[\vec p_1,\vec p_2,m_1,m_2]=\int \frac{d^2k}{(2\pi)^2}\frac{k_{\mu_1}\dots k_{\mu_p}}{\big(\vec k^2-m^2_1\big)\big((\vec k-\vec p_1-\vec p_2)^2-m^2_2\big)}
, \\ \nn 
&&
T_{\mu_1\dots \mu_p}\big[m]=\int \frac{d^2k}{(2\pi)^2}\frac{k_{\mu_1}\dots k_{\mu_p}}{\big(\vec k^2-m^2\big)}
\eea 
where the $B$ and $T$ label denote bubble and tadpole type diagrams. 

\subsubsection*{NFS integrals }
The integrals we encounter for the NFS string are versions of (\ref{eq:standard-integrals}) with pure powers of $k_-$ in the numerator. Furthermore, for processes with massive external legs the two particles in the loop always have the same mass. Using this and denoting the power of $k_-$ with $r$ we can make a slight improvement of the notation as
\bea \nn 
B_r\big[\vec p,\vec p\,',m\big]=\frac{1}{(2\pi)^2}\int d^2k\frac{k_-^r}{\big(\vec k^2-m^2\big)\big((\vec k -\vec p- \vec p\,')^2-m^2\big)}
\eea 
which after introduction of a Feynman parameter becomes
\bea \nn 
B_r\big[\vec p ,\vec p\,',m\big]=
\frac{i}{4\pi}\int dx \frac{\big((1-x)(p_-+p'_-)\big)^r}{m^2+(x-1)x\big(\vec p+\vec p\,'\big)^2}
\eea 
Assuming that $p_1>p'_1$, which implies $p'_->p_-$, and using an IR regulator $\rho$ we find for the relevant integrals
\bea 
&& \textrm{s-channel:} \\ \nn 
&& B_0\big[\vec p,\vec p\,',1\big]=-\frac{1}{2\pi}\frac{p_-p'_-}{(p'_-)^2-p_-^2}\big(i\log\frac{p'_-}{p_-}+\pi\big),\qquad 
B_1\big[\vec p,\vec p\,',1\big]=-\frac{1}{4\pi}\frac{p_-p'_-}{p'_--p_-}\big(i\log\frac{p'_-}{p_-}+\pi\big) , 
\\ \nn 
&& \textrm{u-channel:} \\ \nn 
&& B_1\big[\vec p ,-\vec p\,',0\big]=\frac{i}{4\pi}\frac{p_- p'_-}{p'_--p_-}\log\rho,\qquad 
B_2\big[\vec p ,-\vec p\,',0\big]=-\frac{i}{4\pi}p_- p'_-\big(1+\log\rho\big), \\ \nn 
&& B_3\big[\vec p ,-\vec p\,',0\big]=\frac{i}{8\pi}p_-p'_-(p'_--p_-)\big(3+2\log\rho\big),\qquad 
B_0\big[\vec p ,-\vec p\,',1\big]=\frac{i}{2\pi}\frac{p_- p'_-}{(p'_-)^2-p_-^2}\log\frac{p'_-}{p_-}, \\ \nn 
&& B_1\big[\vec p ,-\vec p\,',1\big]=-\frac{i}{4\pi}\frac{p_- p'_-}{p'_-+p_-}\log\frac{p'_-}{p_-}, \quad 
B_2\big[\vec p ,-\vec p\,',1\big]=-\frac{i}{4\pi}\frac{p_- p'_-\big(\big(1-\log\frac{p'_-}{p_-}\big)(p'_-)^2-\big(1+\log\frac{p'_-}{p_-}\big)p_-^2\Big)}{(p'_-)^2-p_-^2}, 
\eea
We also have t-channel processes. These however turn out to be rather subtle when we have massless particles propagating in the loops. The best way to evaluate them is to perform the $x$ integration using the u-channel integrals and at the end expand around $\vec p\,'\rightarrow \vec p$.

\section{Details for massless scattering}
The one-loop elements for the massless processes $S^{(1)}_{ij}, T^{(1)}_{ij}$ and $R^{(1)}_{ij}$, are encoded in the following expressions\footnote{Expressions like $\big(\vec p_1\pm \vec p_2\big)_-$ etc mean the minus component of the vector $\vec p_1\pm \vec p_2$.}
\bea \label{eq:massless33}
&& S^{(1)}_{33}=2\big(\epsilon_1^2+p_1^2\big)\big(\epsilon_2^2+p_2^2\big)B\big[\vec p_1,-\vec p_1,1\big]-(\vec p_1)^2(\vec p_2)^2B_{+-}\big[\vec p_1,-\vec p_1,0\big] \\ \nn 
&&
+\big(2(\vec p_1\cdot \vec p_2)^2-(\vec p_1)^2(\vec p_2)^2\big)\Big(\big(\vec p_1-\vec p_2)_+B_-\big[\vec p_1,-\vec p_2,0\big]-B_{+-}\big[\vec p_1,-\vec p_2,0\big]\Big) \\ \nn 
&&+2\big(\epsilon_1\epsilon_2+p_1p_2\big)^2 B\big[\vec p_1,-\vec p_2,1\big]  +2(\vec p_1\cdot \vec p_2)^2T\big[1\big], \\ \nn
&& T^{(1)}_{3\bar 3}=2\big(\epsilon_1^2+p_1^2\big)\big(\epsilon_2^2+p_2^2\big)B\big[\vec p_1,-\vec p_1,1\big]-(\vec p_1)^2(\vec p_2)^2B_{+-}\big[\vec p_1,-\vec p_1,0\big] \\ \nn 
&&
+\big(2(\vec p_1\cdot \vec p_2)^2-(\vec p_1)^2(\vec p_2)^2\big)\Big(\big(\vec p_1+\vec p_2)_+B_-\big[\vec p_1,\vec p_2,0\big]-B_{+-}\big[\vec p_1,\vec p_2,0\big]\Big)\\ \nn 
&& +2\big(\epsilon_1\epsilon_2+p_1p_2\big)^2 B\big[\vec p_1,\vec p_2,1\big]  +2(\vec p_1\cdot \vec p_2)^2T\big[1\big], \\ \nn
&& R^{(1)}_{3\bar 3}=2\big(\epsilon_1\epsilon_2+p_1p_2\big)^2 B\big[\vec p_1,\vec p_2,1\big]+(\vec p_1)^2(\vec p_2)^2\Big(\big(\vec p_1+\vec p_2\big)_+ B_-\big[\vec p_1,\vec p_2,0\big]-B_{+-}\big[\vec p_1,\vec p_2,0\big]\Big)
\\ \nn 
&& 
+(\vec p_1)^2(\vec p_2)^2\Big(\big(\vec p_1-\vec p_2\big)_+ B_-\big[\vec p_1,-\vec p_2,0\big]
-B_{+-}\big[\vec p_1,-\vec p_2,0\big]\Big) \\ \nn 
&&+2\big(\epsilon_1\epsilon_2+p_1p_2\big)^2B\big[\vec p_1,-\vec p_2,1\big] +2 (\vec p_1)^2(\vec p_2)^2 T\big[1\big]
\eea 
while the scattering among the second set of massless coordinates gives
\bea \label{eq:massless44}
&& S^{(1)}_{44}=2\big(\epsilon_1^2+p_1^2\big)\big(\epsilon_2^2+p_2^2\big)B\big[ \vec p_1,-\vec p_1,1\big]+2\big(\epsilon_1\epsilon_2+p_1p_2\big)^2B\big[\vec p_1,-\vec p_2,1\big] \\ \nn 
&& +2(\vec p_1\cdot \vec p_2)^2\Big(\big(\vec p_1+\vec p_2\big)_+B_-\big[\vec p_1,\vec p_2,0\big]-B_{+-}\big[\vec p_1,\vec p_2,0\big]\Big)
+2(\vec p_1\cdot \vec p_2)^2 T\big[1\big], \\ \nn 
&& T^{(1)}_{4\bar 4}=2\big(\epsilon_1^2+p_1^2\big)\big(\epsilon_2^2+p_2^2\big)\big[ \vec p_1,-\vec p_1,1\big]+2\big(\epsilon_1\epsilon_2+p_1p_2\big)^2B\big[\vec p_1,\vec p_2,1\big] \\ \nn 
&& +2(\vec p_1\cdot \vec p_2)^2\Big(\big(\vec p_1-\vec p_2\big)_+B_-\big[\vec p_1,-\vec p_2,0\big]-B_{+-}\big[\vec p_1,-\vec p_2,0\big]\Big)
+2(\vec p_1\cdot \vec p_2)^2 T\big[1\big], \\ \nn 
&& R^{(1)}_{4\bar 4}=-2(\vec p_1)^2(\vec p_2)^2B_{+-}\big[\vec p_1,-\vec p_2,0\big]
+2\big(\epsilon_1\epsilon_2+p_1p_2\big)^2\Big(B\big[\vec p_1,\vec p_2,1\big]+B\big[\vec p_1,-\vec p_2,1\big]\Big) \\ \nn 
&& + 2(\vec p_1)^2(\vec p_2)^2 T\big[1\big]
\eea 
For processes mixing different particles we find
\bea \label{eq:massless43}
&& S^{(1)}_{43}=2\big(\epsilon_1^2+p_1^2\big)\big(\epsilon_2^2+p_2^2\big)B\big[\vec p_1,-\vec p_2,1\big] +\big(2(\vec p_1\cdot \vec p_2)^2-(\vec p_1)^2(\vec p_2)^2\big)T\big[1\big]\\ \nn 
&&
+\frac{1}{2}(\epsilon_1-p_1)^2(\epsilon_2+p_2)^2\Big(\big(\vec p_1-\vec p_2\big)_+ B_{-}\big[\vec p_1,-\vec p_2,0\big]-B_{+-}\big[\vec p_1,-\vec p_2,0\big]\Big) \\ \nn 
&&
+\frac{1}{2}(\epsilon_1+p_1)^2(\epsilon_2-p_2)^2\Big(\big(\vec p_1+\vec p_2\big)_+ B_{-}\big[\vec p_1,\vec p_2,0\big]-B_{+-}\big[\vec p_1,\vec p_2,0\big]\Big), \\ \nn 
&& T^{(1)}_{4\bar 3}=2\big(\epsilon_1^2+p_1^2\big)\big(\epsilon_2^2+p_2^2\big)B\big[\vec p_1,-\vec p_2,1\big] +\big(2(\vec p_1\cdot \vec p_2)^2-(\vec p_1)^2(\vec p_2)^2\big)T\big[1\big]\\ \nn 
&&
+\frac{1}{2}(\epsilon_1-p_1)^2(\epsilon_2+p_2)^2\Big(\big(\vec p_1+\vec p_2\big)_+ B_{-}\big[\vec p_1,\vec p_2,0\big]-B_{+-}\big[\vec p_1,\vec p_2,0\big]\Big) \\ \nn 
&&
+\frac{1}{2}(\epsilon_1+p_1)^2(\epsilon_2-p_2)^2\Big(\big(\vec p_1-\vec p_2\big)_+ B_{-}\big[\vec p_1,-\vec p_2,0\big]-B_{+-}\big[\vec p_1,-\vec p_2,0\big]\Big), \\ \nn 
&& R^{(1)}_{4\bar 3}=2\big(\epsilon_1\epsilon_2+p_1p_2\big)^2\big[\vec p_1,\vec p_2,1\big] -\frac{1}{2}(\vec p_1)^2(\vec p_2)^2B_{+-}\big[\vec p_1,-\vec p_1,0\big] \\ \nn 
&& +\frac{1}{2}(\vec p_1)^2(\vec p_2)^2\Big(\big(\vec p_1-\vec p_2\big)_+B_-\big[\vec p_1,-\vec p_2,0\big]-B_{+-}\big[\vec p_1,-\vec p_2,0\big]\Big)+(\vec p_1)^2(\vec p_2)^2T\big[1\big]
\eea 

\section{Details for massive scattering}
Focusing on $y_2$ particles and using (\ref{eq:standard-integrals}) we have,
\bea 
\label{eq:appendix-massive-integrals}
&& S_{22}^{(1)}=\frac{1}{2}\Big[p_- p'_-\Big(p_- p'_- B_0\big[\vec p,-\vec p,1]+B_{2}\big[\vec p,-\vec p,1]-B_{2}\big[\vec p,-\vec p,0]\Big)\Big]_t \\ \nn 
&& +\frac{1}{4}\Big[\big(p_-+p'_-\big)^3\Big(\big(p_-+p'_-\big)B_0\big[\vec p,\vec p\,',1\big]-B_1\big[\vec p,\vec p\,',1\big]\Big)\Big]_s \\ \nn 
&& +\frac{1}{4}\Big[\big(p_-^4+p_-^2(p'_-)^2-p_-(p'_-)^3+(p'_-)^4\big)B_0\big[\vec p,-\vec p\,',1\big]+(p_--p'_-)\big(2p_-^2-p_-p'_-+(p'_-)^2\big)B_1\big[\vec p,-\vec p\,',0\big] \\ \nn 
&& -2\big(p_-^3+p_-(p'_-)^2-(p'_-)^3\big) B_1\big[\vec p,-\vec p\,',1\big]
-\big(4p_-^2-3p_- p'_-+(p'_-)^2\big)B_{2}\big[\vec p,-\vec\, p',0\big] \\ \nn 
&& +\big(p_-^2+(p'_-)^2\big)B_{2}\big[\vec p,-\vec p\,',1\big]\Big]_u, \\ \nn 
&& T_{22}^{(1)}=-\frac{1}{4}\Big[\big(p_--p'_-\big)^3\Big(B_1\big[\vec p,-\vec p\,',1\big]+\big(p'_--p_-\big)B_0\big[\vec p,-\vec p\,',1\big]\Big)\Big]_u+\frac{1}{4}\Big[\big(p_-^2+(p'_-)^2\big)B_2\big[\vec p,\vec p\,',1\big] \\ \nn 
&& -2\big(p_-^3+p_-(p'_-)^2+(p'_-)^3\big)B_1\big[\vec p,\vec p',1\big]
+\big(p_-^4+p_-^2(p'_-)^2+p_-(p'_-)^3+(p'_-)^4\big)B_0\big[\vec p,\vec p\,',1\big] \\ \nn 
&& +2p_-B_3\big[\vec p,\vec p\,',0\big]
-\big(4p_-^2+3p_-p'_-+(p'_-)^2\big)B_2\big[\vec p,\vec p\,',0\big]
+(p_-+p'_-)\big(2p_-^2+p_-p'_-+(p'_-)^2\big)B_1\big[\vec p,\vec p\,',0\big]\Big]_s \\ \nn 
&& +\frac{1}{4}\Big[p_- p'_- \big(B_2\big[\vec p,-\vec p,0\big] +B_2\big[\vec p\,',-\vec p\,',0\big]\big)+(p_- p'_- )^2\big(B_0\big[\vec p,-\vec p,1\big] +B_0\big[\vec p\,',-\vec p\,',1\big]\big)\Big]_t
\eea
where we made use of the trivial fact that t-channel integrals with $r\geq 3$ are zero and $B_r\big[\vec p,-\vec p,1\big]$ for $r=1,2$ also equals zero. For the remaining $33$ and $23$ processes we have similar expressions.

\section{Relevant piece of quartic Lagrangian}
Here we collect the terms in the quartic Lagrangian of the form $(y_2^2,y_2y_3,y_3^2)\times\chi^2$ \footnote{To keep the expression as compact as possible we here denote $\partial_+$ with dot and $\partial_-$ with prime.}
\begin{align}
\label{Lbf}
\mathcal{L}^4_{BF} &= 
%\frac{i}{4}\sum^{4}_{i=1} \Big(\dot{\chi}^i_+\bar{\chi}^i_+ + (\chi^i_-)' \bar{\chi}^i_-\Big)\ |y_1|^2 \\ \nn &\quad 
-\frac{i}{4}\alpha^2\ \left[\left(\dot{\chi}^i_+\bar{\chi}^i_+ + (\chi^i_-)' \bar{\chi}^i_- \right) - 4i\ (1-\alpha)\left(\chi^2_- \bar{\chi}^2_+ - \chi^3_- \bar{\chi}^3_+\right)\right]|y_2|^2
\nonumber\\ 
&\quad - \frac{i}{4}(1-\alpha)^2\left[\left(\dot{\chi}^i_+ \bar{\chi}^i_+ + (\chi^i_-)' \bar{\chi}^i_-\right) + 4i\alpha\left(\chi^2_- \bar{\chi}^2_+ - \chi^3_- \bar{\chi}^3_+\right)\right]|y_3|^2
\nonumber\\
%&\quad - \frac{1}{2}\left(\chi^1_-\bar{\chi}^1_+ + \alpha\ \chi^2_-\bar{\chi}^2_+ + (1-\alpha)\ \chi^3_+\bar{\chi}^3_-\right)\dot{\bar{y}}_1 y'_1\\ \nonumber
%&\quad - \frac{i}{4}\left[\left(\chi^1_-\bar{\chi}^1_- + \chi^2_-\bar{\chi}^2_- - \chi^3_-\bar{\chi}^3_- - \chi^4_-\bar{\chi}^4_-\right)-\left(\chi^1_+\bar{\chi}^1_+ + \chi^2_+\bar{\chi}^2_+ - \chi^3_+\bar{\chi}^3_+ - \chi^4_+\bar{\chi}^4_+\right)\right]y_1(\dot{\bar{y}}_1 - \bar{y}'_1)\\ \nonumber
&\quad + \frac{1}{2}\left(\alpha\chi^1_+\bar{\chi}^1_- +  \chi^2_+\bar{\chi}^2_- + (1-\alpha)\chi^4_+\bar{\chi}^4_-\right)\ \dot{\bar{y}}_2 y'_2 -\frac{i}{4}\alpha\chi^i_-\bar\chi^i_-\,y_2\big(\dot{\bar{y}}_2-\alpha\bar{y}_2'\big)
\nonumber\\ 
&\quad -\frac{i}{4}\alpha\,\chi^i_+\bar\chi^i_+\,y_2\big(\bar{y}_2'-\alpha\dot{\bar{y}}_2\big)
+\frac{1}{2}\left((1-\alpha)\chi^1_+\bar{\chi}^1_- + \chi^3_-\bar{\chi}^3_+ + \alpha\chi^4_+\bar{\chi}^4_-\right)\dot{\bar{y}}_3 y'_3
\nonumber\\ 
&\quad - \frac{i}{4}(1-\alpha)
\left(\chi^1_-\bar{\chi}^1_- - \chi^2_-\bar{\chi}^2_- - \chi^3_-\bar{\chi}^3_- + \chi^4_-\bar{\chi}^4_-\right)\,y_3(\dot{\bar{y}}_3-(1-\alpha)\ \bar{y}'_3)  
\nn \\ 
& \quad -\frac{i}{4}(1-\alpha)\left(\chi^1_+\bar{\chi}^1_+ - \chi^2_+\bar{\chi}^2_+ - \chi^3_+\bar{\chi}^3_+ + \chi^4_+\bar{\chi}^4_+\right)\,y_3(\bar{y}'_3-(1-\alpha)\ \dot{\bar{y}}_3)
 \nn \\ 
&\quad
% + \frac{1}{2}\Big((1-\alpha)\ \chi^2_-\bar{\chi}^2_+ +\alpha\ \chi^3_+\bar{\chi}^3_- + \chi^4_-\bar{\chi}^4_+\Big)\dot{y}_4 y'_4 
%
-\frac{i}{2}\chi_-^3\chi_+^2\,\dot{\bar{y}}_2y_3'
-\frac{i}{2}\chi_-^2\chi_+^3\,\bar y_2'\dot y_3
+\frac{1}{2}\chi^2_+\chi_+^3\left((1-\alpha)^2y_3\dot{\bar{y}}_2-\alpha^2\dot y_3\bar{y}_2\right)
\nn \\ 
&\quad
-\frac{1}{2}\chi^2_-\chi_-^3\left((1-\alpha)^2y_3\bar{y}_2'-\alpha^2y_3'\bar{y}_2\right)
%\\ \nn 
%&\quad 
+\frac{i}{2}\big(\alpha \bar \chi^1_-\bar \chi^4_+-(1-\alpha)\bar \chi_-^4\bar \chi_+^1\big)\dot y_3 y'_2 \\ \nn 
&\quad + 
\frac{i}{2}\big(\alpha\bar \chi_-^4\bar \chi_+^1-(1-\alpha) \bar \chi^1_-\bar \chi^4_+\big)\dot y_2 y'_3
+\frac{1}{2}\big(\alpha^2y'_3 y_2-(1-\alpha)^2y'_2 y_3\big)\bar \chi_-^1 \bar \chi_-^4 \\ \nn 
& \quad 
+\frac{1}{2}\big((1-\alpha)^2\dot y_2 y_3-\alpha^2 \dot y_3 y_2\big)\bar \chi_+^1 \bar \chi_+^4
+ h.c. + ...\ ,
\end{align}
where the ellipses denote parts not relevant for our computations.

\end{appendix}

\end{document}